\documentclass[twocolumn]{aastex62}
\definecolor{emerald}{rgb}{0, 0.5, 0.2}

\newcommand{\ts}{\textsuperscript}

\newcommand{\arxiv}{\textcolor{black}}

\usepackage{soul, color, xcolor,times, aas_macros}
\usepackage{amsmath}
\usepackage{natbib}

\accepted{\today}
\submitjournal{ApJ}

\shorttitle{The Morphology-Density relationship in $1<z<2$ clusters}
\shortauthors{Sazonova et al.}

\begin{document}

\title{The Morphology-Density Relationship in $1<z<2$ clusters}


\author[0000-0001-6245-5121]{Elizaveta Sazonova}
\affiliation{Johns Hopkins University \\
Department of Physics and Astronomy \\
Baltimore, MD 21218, USA}

\author[0000-0002-4261-2326]{Katherine Alatalo}
\affiliation{Space Telescope Science Institute \\
3700 San Martin Dr \\
Baltimore, MD 21218, USA}
\affiliation{Johns Hopkins University \\
Department of Physics and Astronomy \\
Baltimore, MD 21218, USA}

\author[0000-0003-3130-5643]{Jennifer Lotz}
\affiliation{Gemini Observatory \\
950 North Cherry Ave \\
Tucson, Arizona 85726, USA}
\affiliation{Space Telescope Science Institute \\
3700 San Martin Dr \\
Baltimore, MD 21218, USA}

\author[0000-0001-7883-8434]{Kate Rowlands}
\affiliation{Space Telescope Science Institute \\
3700 San Martin Dr \\
Baltimore, MD 21218, USA}
\affiliation{Johns Hopkins University \\
Department of Physics and Astronomy \\
Baltimore, MD 21218, USA}

\author[0000-0002-4226-304X]{Gregory F. Snyder}
\affiliation{Space Telescope Science Institute \\
3700 San Martin Dr \\
Baltimore, MD 21218, USA}

\author[0000-0002-5828-6211]{Kyle Boone}
\affiliation{Department of Physics\\
University of California Berkeley\\
Berkeley, CA 94720, USA}

\author[0000-0002-4208-798X]{Mark Brodwin}
\affiliation{Department of Physics and Astronomy\\
University of Missouri\\
Kansas City, MO 64110, USA}

\author[0000-0001-9200-8699]{Brian Hayden}
\affiliation{Department of Physics\\
University of California Berkeley\\
Berkeley, CA 94720, USA}

\author[0000-0002-3249-8224]{Lauranne Lanz}
\affiliation{The College of New Jersey\\
Ewing, NJ 08618, USA}

\author[0000-0002-4436-4661]{Saul Perlmutter}
\affiliation{Department of Physics\\
University of California Berkeley\\
Berkeley, CA 94720, USA}

\author[0000-0002-9495-0079]{Vicente Rodriguez-Gomez}
\affiliation{Instituto de Radioastronom\'ia y Astrof\'isica\\
Universidad Nacional Aut\'onoma de M\'exico \\
Apdo. Postal 72-3, 58089 Morelia, Mexico}

\begin{abstract}

The morphology-density relationship states that dense cosmic environments such as galaxy clusters have an overabundance of quiescent elliptical galaxies, but it is unclear at which redshift this relationship is first established. We study the morphology of 4 clusters with 1.2$<$z$<$1.8 using \textit{HST} imaging and the morphology computation code \textsc{statmorph}. By comparing median morphology of cluster galaxies to CANDELS field galaxies using Monte Carlo analysis, we find that 2 out of 4 clusters (at z=1.19 and z=1.75) have an established morphology-density relationship with more than $3\sigma$ significance. $\sim$50\% of galaxies in these clusters are bulge-dominated compared to $\sim$30\% in the field, and they are significantly more compact. This result is more significant for low-mass galaxies with $\log M/M_\odot \lessapprox 10.5$, showing that low-mass galaxies are affected the most in clusters. We also find an intriguing system of two z$\approx$1.45 clusters at a unusually small separation 2D separation of $3'$ and 3D separation of $\approx$73 Mpc that exhibit no morphology-density relationship but have enhanced merger signatures. We conclude that the environmental mechanism responsible for the morphology-density relationship is 1) already active as early as z=1.75, 2) forms compact, bulge-dominated galaxies and 3) affects primarily low-mass galaxies. However, there is a significant degree of intracluster variance that may depend on the larger cosmological environment in which the cluster is embedded.
\end{abstract}

 
\section{Introduction} \label{sec:intro}

In the local Universe, the environment of a galaxy plays a crucial role in its evolution. This has become known as the morphology-density relationship: galaxy clusters, the densest environments, are populated primarily by early-type galaxies, while spiral galaxies are less common there \citep{Dressler1980}. Morphology of a galaxy is strongly correlated with its star formation history \citep[e.g.,][]{Larson1980,Strateva2001,Lee2013}, and the vast majority of cluster galaxies are  quiescent.

However, it is important to understand the role of environment at earlier times. The period of $1<z<3$ is crucial for galaxy evolution, as high-redshift galaxies undergo dramatic changes to form their present-day descendants. At $z\sim3$, massive galaxies are still predominantly star-forming disks. Star formation rate (SFR) peaks at $1<z<2$ and then steadily drops across all populations of galaxies \citep{Madau2014}. During the $1<z<2$ period, the fraction of massive quiescent galaxies rapidly grows, reaching 80\% at $z=1$ \citep{Muzzin2013, Buitrago2013}. At $z>1.5$, this population consists primarily of compact quiescent galaxies (cQGs), which then grow to form larger early-type galaxies (ETGs) we see today \citep{Cassata2011,Cassata2013,vanDokkum2015,vanderWel2014}. Multiple formation mechanisms have been proposed for cQGs, including mergers \citep[e.g.,][]{Hopkins2009,Wuyts2010,Wellons2015} and violent disk instabilities \citep[VDIs; e.g.,][]{Dekel2013,Zolotov2015}. It is also during this period that the cosmic web collapses sufficiently to form dense clusters for the first time \citep{Muldrew2015}, and morphology-density relationship must begin to establish.

Multiple environmental quenching mechanisms have been proposed and observed in the local Universe, such as ram-pressure stripping \citep{Gunn1972}, harassment \citep{Moore1996}, strangulation \citep{Larson1980} and group pre-processing \citep{Wilman2008,Dressler2013}. In addition, there is evidence of an enhanced merger rate in high-redshift clusters \citep{Lotz2013,Watson2019}, so quenching via mergers could also be more important in these systems. 

Yet, the influence of the environment on galaxy properties at $1<z<3$ is still highly disputed. Observational studies of star formation in high-redshift clusters are few and often conflicting. Multiple recent studies find a factor of 2-3 suppression in star formation in $z<2$ cluster galaxies \citep{Quadri2012,Cooke2016,Lee-Brown2017,Kawinwanichakij2017,Strazzullo2019,vanderBurg2020}, while \cite{Tran2010} find a reversal of SFR-density relation at $z=1.62$. \cite{Papovich2012} find that cluster galaxies are, on average, larger than in the field, while \cite{Newman2014} find no significant difference. Finally, \cite{Brodwin2013} note a large degree of scatter from cluster to cluster, making any generalizations difficult. A major problem in making a statistical statement on environmental influence is still a lack of data. Galaxy clusters are only beginning to form at $z > 1$, they are difficult to identify and to distinguish from protoclusters \citep{Cautun2014,Muldrew2015}. Moreover, establishing cluster membership is challenging at high redshifts. Finally, reliable morphology metrics are difficult to compute at these redshifts, since high resolution imaging is required. 

With the advent of space-based telescopes alongside increasing computational power, it has become easier to probe higher redshifts at better spatial resolution. A big breakthrough was the development of numerical morphology estimates such as S\'ersic Index \citep{Sersic1963}, and consequent development of non-parametric statistics, such as Concentration, Asymmetry, Smoothness \citep{Conselice2003} and Gini/$M_{20}$ \citep{Lotz2004}. Numerical measurements have made statistical analysis of large samples of high-redshift galaxies much more computationally feasible. A large (120 sq. arcmin) area of the sky was observed with the \textit{Hubble Space Telescope} (HST) in the CANDELS survey \citep{candels1,candels2}, and non-parametric morphology parameters were computed for CANDELS galaxies first by \cite{Peth2016} and consequently by Rodriguez-Gomez (in prep.), using an open-source code \textsc{statmorph} \citep{statmorph}. However, morphological studies of cluster galaxies are still few and often focus on a smaller subset of traditional morphology parameters, such as radius and S\'ersic index \citep[e.g.,][]{Papovich2012,Buitrago2013,Newman2014,Socolovsky2019}. In this work, we extend the analysis of morphology of 1$<$z$<$2 cluster galaxies to a wide range of structural parameters, including Concentration, Gini and $M_{20}$.

We study 4 clusters spectroscopically identified by \cite{Brodwin2013} and references therein, and use public HST data to probe the morphology of the cluster galaxies. We show evidence for the existence of the morphology-density relationship as early as z$\sim$1.75, but also a significant variation from cluster to cluster. This variation highlights differences in the formation process of individual clusters, and upon further study may shed light on the mechanisms of structure formation, as well as the timescale over which the morphology-density relationship is established.

In Section \ref{sec:data}, we describe the cluster and the field samples. In Section \ref{sec:analysis}, we describe the morphological and statistical analysis performed in this study, including new bulge strength and disturbance metrics derived in this paper using Principal Component Analysis (PCA). Results of statistical comparison of cluster galaxies to a control sample of CANDELS field galaxies are shown in Section \ref{sec:results} and discussed in Section \ref{sec:discussion}. Finally, we present our conclusions in Section \ref{sec:conclusions}. Throughout this work, to maintain consistency with photometric redshifts from \cite{Brodwin2013}, we use \textit{Wilkinson Microwave Anisotropy Probe 7} (WMAP 7) cosmology with $(\Omega_\Lambda, \Omega_M, h) = (0.728, 0.272, 0.704)$ \citep{wmap7}. 

\arxiv{\vfill}
\section{Data}\label{sec:data}

\begin{deluxetable*}{rccccccclccc}[hbt!]
\tablecaption{Clusters used in this paper}
\tablehead{
\colhead{Cluster} & \colhead{z} & \colhead{$N_{\textrm{phot+spec}}$\tablenotemark{(1)}} & \colhead{$N_{\textrm{spec}}$\tablenotemark{(2)}} &
\colhead{$N_{\textrm{HST}}$\tablenotemark{(3)}} & \colhead{$N_{\textrm{flag}}$\tablenotemark{(4)}} &
\colhead{$M_{14}$\tablenotemark{(5)}} & \colhead{$R_{200}$} &
\colhead{Filter}  & \colhead{$t_{\textrm{exp}}$} & \colhead{SB Limit} & \colhead{Ref.} \\
 &  &  & & & & & \colhead{(kpc)} & &
\colhead{(s)} & \colhead{(mag/arcsec$^2$)} & 
}
\startdata
J1142+1527 & 1.19 & 121 & 5 & 107 & 4 & $11\pm2$ & $1400\pm 80$ & F140W & 4988 & 26.1 & G15\\
J1432.3+3253 & 1.40 & 172 & 19 & 30 & 0 &  -- & -- & F140W & 11224 & 26.5 & B13, Z13\\
J1432.4+3250 & 1.49 & 159 & 6 & 143 & 16 & $2.5\pm0.5$ & $760\pm50$ & F160W\tablenotemark{(6)}
& 2611 & 25.4 & B11, B13\\
J1426.5+3508 & 1.75 & 46 & 7 & 45 & 1 & $4.5\pm0.5$ & $710\pm45$ & F160W\tablenotemark{(6)} 
& 7558 & 25.8 & S12, B16 \\
\enddata 
\noindent \tablenotemark{(1)}{Number of cluster members identified in other works.} \\
\tablenotemark{(2)}{Number of spectrosopic members} \\
\tablenotemark{(3)}{Number of members in the HST imaging} \\
\tablenotemark{(4)}{Number of members flagged by \textsc{statmorph} as having unreliable morphology}\\
\tablenotemark{(5)}{Total cluster mass $M/M_\odot = 10^{14} M_{14}$ \quad} \\
\tablenotemark{(6)}{Drizzled \textit{HST} images obtained from the \textit{Hubble Legacy Archive}}
\tablecomments{G15: \cite{Gonzalez2015}, B11: \cite{Brodwin2011}, B13: \cite{Brodwin2013}, B16: \cite{Brodwin2016}, Z13: \cite{Zeimann2013}, S12: \cite{Stanford2012}}
\label{tab:clusters}\end{deluxetable*}

\begin{figure*}[htb!]
\centering
\includegraphics[width=\textwidth]{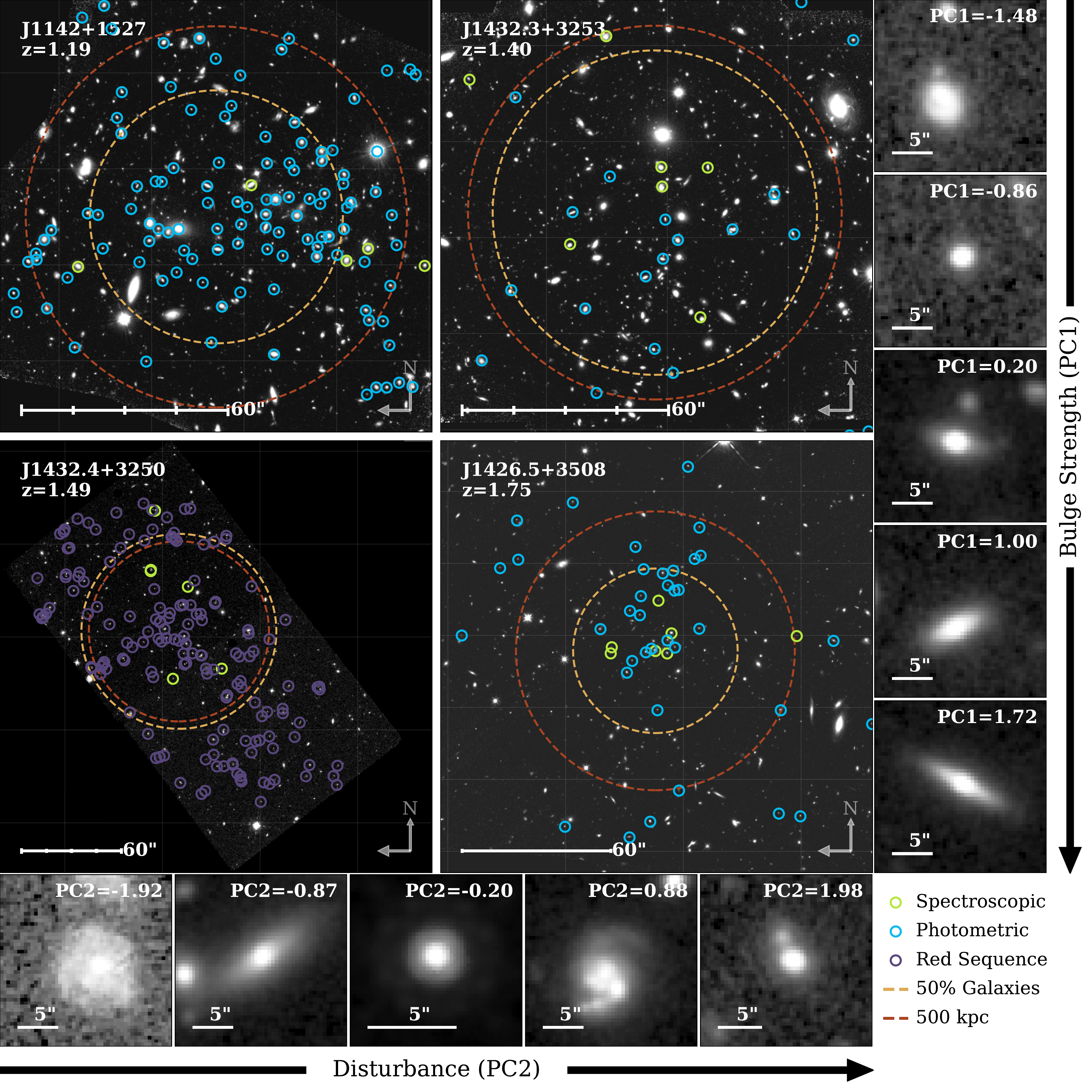}
\caption{\textbf{Center:} HST images of all 4 clusters outlined in Table \ref{tab:clusters}, in a corresponding filter for each cluster (F140W/F160W). Positions of cluster members are shown using photometric (blue) or red-sequence (red) catalogs. Spectroscopically confirmed cluster members are in green. \textbf{Right:} a sample of galaxies in the order of increasing bulge strength (PC1; see Sec. \ref{sec:pca}). \textbf{Bottom:} a sample of galaxies in the order of increasing disturbance (PC2; see Sec. \ref{sec:pca}). Samples of different PC1/2 combinations are shown in Appendix \ref{app:pca}.}
\label{fig:ds9}
\end{figure*}

We analyzed 4 spectroscopically confirmed galaxy clusters at various redshifts: J1432.4+3253 (z$=$1.49), J1432.3+3253 (z$=$1.40), J1142+1527 (z$=$1.19) and J1426.5+3508 (z$=$1.75). These clusters were identified as a part of various programs:  IRAC Shallow Cluster Survey \citep[ISCS;][]{Eisenhardt2008}, its extension, IRAC Deep Cluster Survey (IDCS) and Massive and Distant Clusters of WISE Survey \citep[MaDCoWS;][]{Gonzalez2019}. Table \ref{tab:clusters} provides a brief summary of all the clusters used in this paper, and the Hubble Space Telescope (HST) Wide-Field Camera 3 (WFC3) images of these clusters are shown on Figure \ref{fig:ds9}. 

For the analysis in this paper, we used HST WFC3 imaging in F140W (central wavelength $\lambda_p = 1392.3$ nm) or F160W ($\lambda_p = 1536.9$ nm) bands, choosing the filter with the longest exposure time for each cluster. Cluster membership was identified using photometric, spectroscopic or red-sequence membership selection. Photometric members were identified using optical $B_{W}RI$ data from NOAO Deep Wide-Field Survey \citep[NDWFS;][]{ndwfs},  3.6 $\mu$m and 4.5 $\mu$m IRAC data from \textit{Spitzer} Deep Wide-Field Survey \citep[SDWFS;][]{sdwfs}, and 24 $\mu$m imaging with Multiband Imaging Photometer for \textit{Spitzer} (MIPS). Then, galaxy SED templates from \cite{Polletta2007} were fit to determine photometric redshifts. The resulting accuracy of the photometric redshifts for our spectroscopically confirmed clusters is $\sigma/(1+z)=0.039$ \citep{Brodwin2013}. Spectroscopic members were identified using Keck multi-object spectroscopy and WFC3 grism spectroscopy. The resulting spectroscopic and photometric catalogs were obtained from \cite{Brodwin2013} and references therein. Red-sequence members for the J1432.4+3250 cluster were found by selecting galaxies along the color-magnitude red sequence using \textit{HST} data in \cite{Snyder2012}. We note that the uncertainties in photometric and especially red-sequence membership estimation mean that some cluster galaxies might be field interlopers, so the significance levels in our comparison with the pure field sample are likely underestimated.

\subsection{Individual Clusters}
\subsubsection{J1142+1527}

J1142+1527 (for simplicity, J1142) is a massive $z=1.19$ cluster discovered as part of the MaDCoWS survey. A detailed study of this cluster is presented in  \cite{Gonzalez2015}. Keck and \textit{Gemini} spectroscopy were used to identify the spectroscopic members of this cluster, and \textit{Spitzer} and NDWFS data were used to identify photometric members. The cluster is robustly detected using the Sunyaev-Zel'dovich (SZ) effect using Combined Array for Research in Millimiter-wave Astronomy (CARMA)\footnote{\href{https://www.mmarray.org/}{https://www.mmarray.org/}}, and its mass is estimated from the SZ effect to be $M_{200} \sim (1.1 \pm 0.2) \times 10^{15} M_\odot$, which currently makes it the most massive confirmed cluster at $z \geq 1.15$. The cluster radius is then $R_{200} = 1400 \pm 80$ kpc.

We used reduced deep F140W HST imaging for this cluster, obtained as part of the Supernova Cosmology Project (PI: Perlmutter). The image is $2.2'\times 2.1'$ in size centered on the cluster, so the cluster is imaged up to $0.38 R_{200}$. The exact center is defined by the centroid of the SZ decrement as $(\alpha, \delta)=(\textrm{11:42:46.6, +15:27:15})$ \citep{Gonzalez2015}. There are 5 spectroscopic cluster members and 98 photometric members inside the HST image. 50\% of the cluster members are contained within 0.66$'$ (330 kpc, or $0.24R_{200}$) from the center.

\subsubsection{J1432.3+3253}

J1432.3+3253 (for simplicity, J1432.3) is a $z=1.40$ cluster observed as part of the ISCS survey. It was first presented by \cite{Zeimann2013} and additional members were published in \cite{Brodwin2013}. There are no X-ray or SZ detections of this cluster, so its mass is unknown. Interestingly, this cluster is only separated by 2$'$ from the J1432.4+3250 cluster, and they are located at similar redshifts, $z=1.40$ and $z=1.49$; therefore, the cluster members from \textit{Spitzer} imaging are mixed between them, and there is a possibility of dynamical interactions between these two clusters.

We used reduced F140W HST imaging from the Supernova Cosmology Project for this cluster. The image is $ 2.4'\times2.1'$ in size, centered around the cluster center, so a roughly $600$ kpc radius of the cluster is imaged. The exact cluster center is $(\alpha, \delta) = (\textrm{14:32:18.31, +32:53:07.8})$ defined during the cluster detection using wavelet decomposition \citep{Eisenhardt2008},
There are 10 spectroscopic members in the HST image. Although the catalog in \cite{Brodwin2013} provides a large number of cluster members, only 20 photometric and 10 spectroscopic members are contained in the HST image,  50\% of which are within 0.84$'$ or 440 kpc from the center. 

\subsubsection{J1432.4+3250}

J1432.4+3250 (for simplicity, J11432.4) is a $z=1.49$ ISCS cluster. We used a red-sequence selected membership catalog from \cite{Snyder2012} and spectroscopic membership presented in \cite{Brodwin2013, Brodwin2011}. \textit{Chandra} X-ray imaging of the hot intracluster medium (ICM) results in a mass estimate of $M_{200} = (2.5 \pm 0.5) \times 10^{14} M_\odot$ \citep{Brodwin2011} and radius $R_{200} = 760 \pm 50$ kpc.

We used reduced F160W HST imaging from HST GO 11663 \citep[PI Brodwin;][]{Snyder2012} obtained from the \textit{Hubble Legacy Archive} (HLA). The HST image is $4.3'\times2.3'$ in size and contains 5 spectroscopic and 122 red sequence selected members. In physical units, the image covers $1.4R_{200} \times 0.8R_{200}$, so the cluster is imaged up to $0.4R_{200}$ in both directions. The cluster center is defined using the wavelet detection centroid, at $(\alpha,\delta)=(\textrm{14:32:24.16, +32:50:03.7})$. Since the image is almost twice as large in one dimension, the galaxies away from the cluster center are sampled better along that direction. 50\% of the HST-imaged cluster members are contained within 1.05$'$ or $0.7R_{200}$, due to the elongation of the image along one direction.

\subsubsection{J1426.5+3508}

J1426.5+3508 (for simplicity, J1426.5) is a $z=1.75$ IR-selected cluster from the IDCS survey. It is one of the most distant clusters found. The spectroscopic and photometric membership catalogs are first presented in \cite{Stanford2012}. \textit{Chandra} X-ray observations, weak and strong lensing, and SZ effect measurements produced consistent, but slightly different mass estimates. Due to its lowest uncertainty, we used X-ray mass $M_{500} = (2.8\pm0.3) \times 10^{14} M_\odot$ \citep{Brodwin2016}. Using the conversion factor of $1.6$ \citep{Duffy2008}, this results in $M_{200} = (4.5 \pm 0.5) \times 10^{14} M_\odot$ and corresponding $R_{200} = 710 \pm 45$ kpc. This high mass makes J1426.5 the most massive found at $z \geq 1.5$.

We used reduced F160W HST imaging from HST GO 11663 \citep[PI: Brodwin;][]{Stanford2012} obtained from the HLA. The HST image is $3.6' \times 3.5' $, therefore covering up tp $1.3 R_{200}$ of the cluster. The cluster center is chosen using the position of the Brightest Cluster Galaxy (BCG), $(\alpha, \delta)=(\textrm{14:26:32.95, +35:08:23.6})$ \citep{Brodwin2013}. The center position derived from the SZ centroid and the X-ray peak are compatible with the BCG position. The image contains 6 spectroscopic and 38 photometric members, 50\% of which are contained within 0.57$'$ or $0.3R_{200}$ from the cluster center.

\subsection{Control Sample}

In our analysis, we used control samples consisting of galaxies from all CANDELS fields \citep{candels1, candels2}. Redshifts and stellar masses for CANDELS galaxies were estimated in \cite{Dahlen2013}. Se\'rsic fits of CANDELS galaxies were previously performed by \cite{vanderWel2014}, and the full suite of morphological measurements was computed by Rodriguez-Gomez (in prep.) using F160W imaging.

In our analysis, we only used CANDELS galaxies with a high F160W signal-to-noise ratio ($\textrm{SNR}>3$). The exposure time and image depth varies across all CANDELS images and fields, with $1\sigma$ sensitivity on average between 25.5-26.5 mag/arcsec$^2$.

Although the clusters were imaged in either F160W or F140W bands, we compared all cluster galaxies to F160W CANDELS imaging, since the CANDELS field was never imaged in F140W. However, F160W and F140W filters are highly overlapping: F140W is centered around $1392.3$ nm with a width of $384$ nm, and F160W is centered around $1536.9$ nm with a width of $268.3$ nm, so F140W almost entirely covers the F160W band. At the average cluster redshift $z\approx1.5$, these bands image $556 \pm 154$ nm and $615 \pm 107$ nm respectively. Therefore, we do not expect a significant morphological difference between these two bands.

For each cluster, we selected a Monte Carlo ensemble of control samples to perform statistical analysis, as described in detail in Sec. \ref{sec:mcmc}.

\section{Data Analysis}\label{sec:analysis}

\subsection{Source Extraction}

\begin{table}[h!]
\centering
\caption{\textsc{SExtractor} parameters used for identifying  galaxies} \label{tab:sextractor}
\begin{tabular}{c|c|l}
\hline
\hline
Parameter & Description &Value\\
\hline
\texttt{DETECT$\_$MINAREA} & Min. galaxy area & 10 px \\
\texttt{DETECT$\_$THRESH} & Min. detection S/N & 0.75 \\
\texttt{FILTER$\_$NAME} & Convolution filter & 9x9px tophat \\
\texttt{DEBLEND$\_$NTHRESH} & Num. of deblending levels & 16 \\
\texttt{DEBLEND$\_$MINCONT} & Min. deblending contrast & 0.0001 \\
\texttt{BACK$\_$SIZE} & Background mesh size & 64 px \\
\texttt{BACK$\_$FILTERSIZE} & Background filter size & 3 \\
\hline
\end{tabular}
\end{table}

First, individual galaxies were identified in the cluster HST images using \textsc{SExtractor} \citep{Bertin1996}.  

The segmentation maps for the CANDELS field were previously computed by Lotz et al. (in prep.) using parameters given in Table  \ref{tab:sextractor}, and the same parameters are used to identify cluster galaxies so that the calculated morphology parameters are consistent.

\begin{figure*}[!t]
\centering
\includegraphics[width=\linewidth]{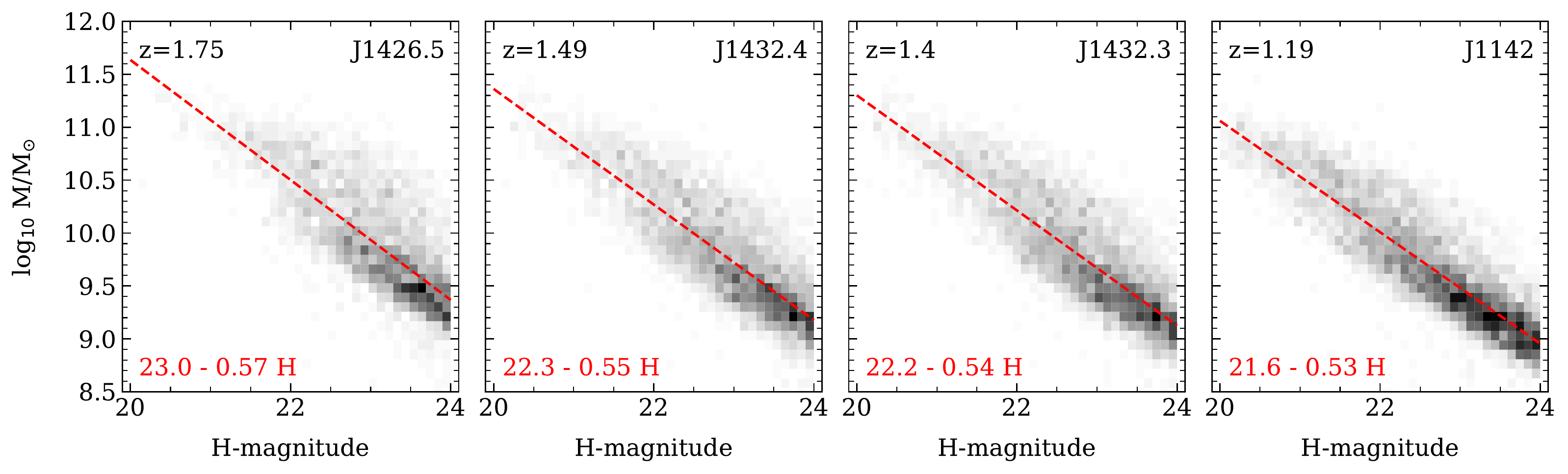}
\caption{The distribution of stellar masses and H-magnitudes for CANDELS field galaxies in four different redshift bins, matching the cluster redshifts with $\Delta z=0.25$. The tight relationship between the H-magnitude and the stellar mass allows us to use H-magnitude as a proxy for mass. The red line shows the line of best fit used to convert H-magnitudes into stellar masses.}
\label{fig:hmag_mass}
\end{figure*}

The tophat filter smoothes the image prior to source extraction and reduces the effect of image noise. Since this work studied small and faint high-redshift galaxies, the detection S/N and minimum galaxy area are set low to allow identifying the faintest cluster members. The deblending level was selected as to allow deblending of merging faint sources.

Magnitudes were also computed at this step using the \texttt{MAG\_AUTO} setting, which calculates flux within an automatically calculated Kron radius \citep{Kron1980}.

\subsection{Morphology}\label{sec:morph}

Morphology of the cluster galaxies was measured using the new code, \textsc{statmorph}  \citep{statmorph}, an open-source Python package for calculating numerical morphology parameters that is largely based on the IDL code described in \cite{Lotz2004, Lotz2006, Lotz2008}. Morphology of the control sample, CANDELS field galaxies, was previously measured in F160W band by Rodriguez-Gomez (in prep.) and following the same methodology as in this paper.

We used a variety of non-parametric statistics to quantify the galaxy morphology: Concentration \citep{Abraham1994, Bershady2000, Conselice2003}, Asymmetry \citep{Schade1995, Abraham1996, Bershady2000, Conselice2003}, Gini \citep{Abraham2003, Lotz2004} and $M_{20}$ \citep{Lotz2004}, all described in detail below. In addition, we used S\'ersic index \citep{Sersic1963} and S\'ersic effective radius. Finally, we performed a PCA to derive a new bulge strength and galaxy disturbance metric, as outlined in Section \ref{sec:pca}.

\subsubsection{S\'ersic Index \& Effective Radius}

S\'ersic index \citeyearpar{Sersic1963} is obtained by fitting the galaxy's light distribution with the S\'ersic profile:

\begin{equation}
    I(R) = I_e \exp \Bigg\{ -b \Bigg[ \bigg( \frac{R}{R_e} \bigg)^{1/n} - 1\Bigg] \Bigg\}
\end{equation}

\noindent where $R_e$ is the effective half-light radius, $I_e$ is the intensity at $R_e$, $n$ is the S\'ersic index, and $b$ is a function of $n$ computed via Gamma functions. At $z=0$, elliptical galaxies are defined as those that follow a de Vaucouleurs \citeyearpar{deVaucouleurs1948} profile with $n=4$, and spiral galaxies follow an exponential profile with $n=1$ \citep{Freeman1970}. \cite{Buitrago2013} show that although S\'ersic index is in general lower for all galaxy types at higher redshifts, a $n=2.5$ limit still separates the two populations well up to $z\sim 2.5$. 

\textsc{statmorph} allows a user to improve the accuracy of the S\'ersic fit by specifying a Point Spread Function (PSF) of the image.  We used a hybrid F160W PSF created by \cite{VanderWel2012} from a \textsc{TinyTim} model \citep{Krist2011} and an empirical fit to CANDELS sources with a $0.06'$ pixel scale. This PSF was then extrapolated to match the pixel scale of each individual galaxy image. The same PSF was used for F140W filter as for F160W filter, since the two bands have similar PSF full width at half maximum.

\subsubsection{Compactness}\label{sec:compactness}

Comparing S\'ersic sizes of galaxies can be misleading, as the galaxy radius is a strong function of its mass. Therefore, we use another metric, Compactness, to compare galaxy sizes normalized by their mass. Compactness is defined similarly to a galaxy's surface mass density, $\Sigma$:

\begin{equation*}
    \Sigma = \frac{M}{\pi R_e^2}
\end{equation*}

where $M$ is the galaxy mass and $R_e$ is its S\'ersic radius. 
Since we do not have stellar masses of cluster galaxies, we use H-band magnitudes as a proxy for masses in the CANDELS field. A correlation between H-band magnitude and stellar mass has been previously shown by \cite{vanderWel2014}. In this work, we checked the correlation by plotting the distribution of H-magnitudes and stellar masses for the CANDELS galaxies in the four required redshift bins.  Fig. \ref{fig:hmag_mass} shows that the tight relationship exists for our field galaxies, which allows us to use H-magnitude as a proxy. To convert from H-magnitude to stellar mass, we fit a line of the form

\begin{equation*}
    \log_{10} \; M/M_\odot = a + bH
\end{equation*}

where $a$ and $b$ are best fit coefficients, shown on Fig. \ref{fig:hmag_mass}. We then define compactness, $\log \Sigma$, as a logarithm of surface mass density:

\begin{equation}
    \log \Sigma = K + bH - 2\log R_e
\end{equation}

where $K = a + \log 2\pi$, a constant absorbing other  coefficients. Note that since $\log \Sigma$ is defined logarithmically and using H-magnitude rather than stellar mass, it will differ from similar measures in other works by some constant factor. However, it is a useful statistic in comparing relative compactness of cluster and field galaxies in our samples.

\subsubsection{Concentration}

Concentration of a galaxy's light (C) has been measured for a long time by various methods \citep[e.g.,][]{Abraham1994}. We used the currently most common definition by \cite{Bershady2000}, where C is is the ratio of the 80\ts{th} and 20\ts{th} isophotes, calculated as 

\begin{equation}
    C = 5 \log_{10} \bigg( \frac{r_{80}}{r_{20}} \bigg)
\end{equation}

Higher values of $C$ mean the light is concentrated in the center, and corresponds to higher S\'ersic index, more bulge dominated and spheroidal galaxies. 

\subsubsection{Asymmetry}

Asymmetry (A) of a galaxy's light distribution is computed by rotating the galaxy by 180$^\circ$ and subtracting the resulting light distribution from the original one \citep{Schade1995, Abraham1996, Conselice2003}. The axis of rotation is chosen as to minimize the asymmetry. 

\begin{equation}
    A = \frac{\sum_{i,j} |I_{ij} - I_{ij}^{180}|}{\sum_{i,j}|I_{ij}|} - A_{BG}
\end{equation}

\noindent where $I_{ij}$ is the flux in a pixel, $I^{180}_{ij}$ is the flux in the corresponding pixel after the rotation and $A_{BG}$ is the average asymmetry of the background. Bulge-dominated galaxies have very low asymmetry values, spiral galaxies typically have intermediate asymmetry and merging galaxies can have very large asymmetry. 
\subsubsection{Gini Index}\label{sec:gini}

Gini index \citep[G;][]{Abraham2003, Lotz2004} originates from economics, where it is used to measure the distribution of wealth in a society. In astronomy, it is used to measure the concentration of light, similar to $C$. To calculate Gini, all of the galaxy pixels are ranked from brightest to dimmest, and Gini is computed as:

\begin{equation}
   G = \frac{1}{\bar X N(N-1)} \sum_{i=0}^N (2i-N-1) X_i
   \label{gini_ms}
\end{equation}

\noindent where $N$ is the number of pixels, $X_i$ is the flux of i\ts{th} pixel ($i=0$ is the brightest pixel in the galaxy) and $\bar{X}$ is the average flux. 

\begin{figure*}[t!]
\centering
\includegraphics[width=1.0\textwidth]{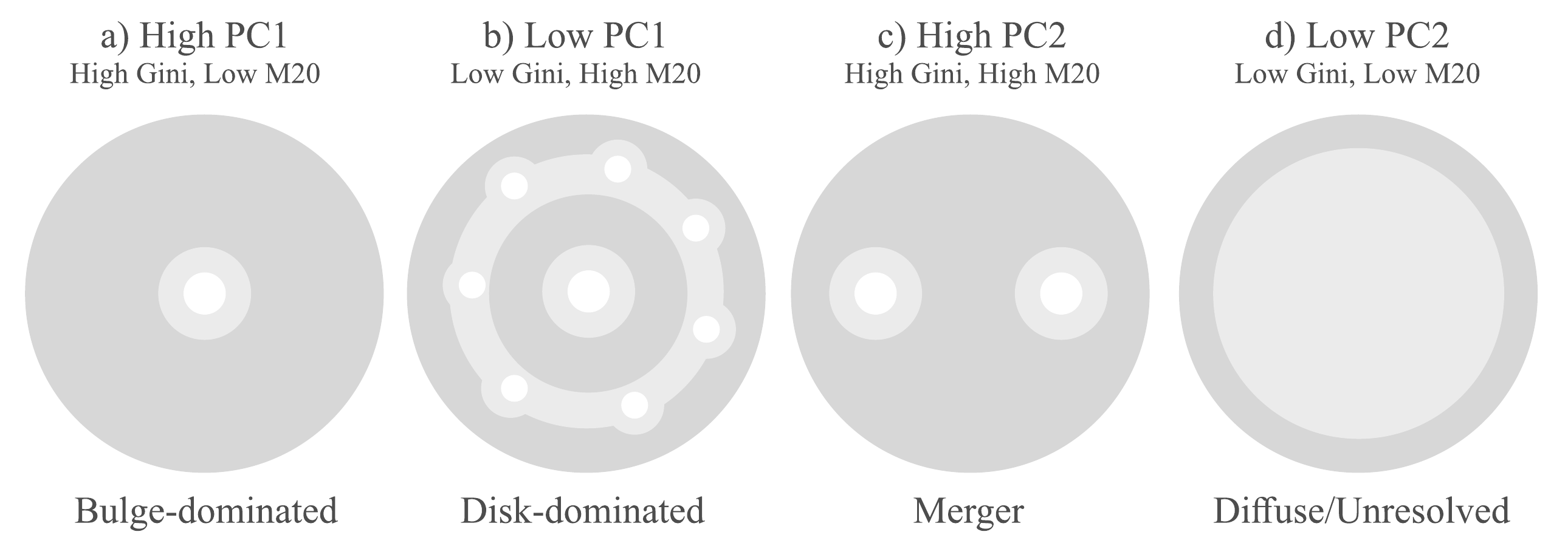}
\caption{A schematic showing the typical galaxy morphologies as measured using the Gini (Sec.\ref{sec:gini}) and $M_{20}$ (Sec.\ref{sec:m20}) non-parametric statistics \citep{Lotz2004}, as well as using the Principal Components presented in Sec.\ref{sec:pca}. \textbf{a)} Bulge-dominated galaxies have high Gini and low $M_{20}$ (high PC1), \textbf{b)} disk-dominated galaxies have low Gini and high $M_{20}$ (low PC1), and \textbf{c)} merging or disturbed galaxies have high Gini and $M_{20}$ (high PC2). \textbf{d)} Galaxies with low Gini and low $M_{20}$  (low PC2) are diffuse and/or poorly resolved.} 
\label{fig:gini_m20}
\end{figure*}

 A Gini index of $1$ signifies all the galaxy light concentrated in one pixel, while a Gini index of $0$  signifies all the light evenly distributed across all pixels that belong to the galaxy. Figure \ref{fig:gini_m20} shows a schematic representation of the Gini index measurement for a galaxy. Highly concentrated spheroidal galaxies have higher $G$ (Fig. \ref{fig:gini_m20}a), while diffuse galaxies, or disky galaxies with extended regions of star formation and weak bulges have low $G$ (Fig. \ref{fig:gini_m20}b). 

The difference between Gini and $C$ is that Gini does not require the brightest pixel to be in the spatial center of the galaxy, so a galaxy with an offset nucleus would still have a high Gini index. Therefore, a combination of $C$ and Gini can potentially identify disturbed morphologies, such as galaxy mergers and double nuclei. In addition, since Gini is computed based on Petrosian radius of the galaxy, it depends very weakly on the limiting fluxes and S/N of the image as long as average S/N per pixel $>$ 2 \citep{Lotz2004}.

\subsubsection{$M_{20}$}\label{sec:m20}

$M_{20}$ is a measure of the second moment of the distribution of the brightest 20\% of galaxy light. It was first introduced in \cite{Lotz2004} as a metric which, together with Gini, can effectively detect signs of ongoing galaxy mergers. The second moment of i\ts{th} pixel is calculated as:

\begin{equation}
    M_i = f_i \cdot [(x_i - x_c)^2 + (y_i - y_c)^2]
\end{equation}

\noindent where $(x_c, y_c)$ are the coordinates of the central pixel and $(x_i, y_i)$, $f_i$ are respectively the coordinates and flux of the i\ts{th} pixel. Then, the $M_{20}$ is calculated using:
\begin{eqnarray}
    M_{\textrm{tot}}  = \sum_{i=0}^N M_i \\
    M_{20} = \log_{10} \sum_{i=0}^{I_{20}} \frac{M_i}{M_{\textrm{tot}}}
\end{eqnarray}
\\

\noindent where $N$ is the total number of pixels and $0 < i < I_{20}$ are the brightest $20\%$ of the pixels. A low (more negative) $M_{20}$ signifies that the galaxy light is concentrated in the center, and corresponds to spheroidal galaxies, as shown on Fig. \ref{fig:gini_m20}a) and d). A high $M_{20}$ means the brightest galaxy light is offset from the center, and corresponds to disk galaxies with bright regions of star formation (Fig. \ref{fig:gini_m20}b) or mergers (Fig. \ref{fig:gini_m20}d). Similarly to Gini, $M_{20}$ calculation is based on the Petrosian radius and hence weakly depends on the limiting fluxes or image S/N \citep{Lotz2004}.

Gini and $M_{20}$ can be used in tandem to detect possible mergers. A high $G$ and a high $M_{20}$ corresponds to a scenario where the galaxy light is concentrated in a few bright spots that are off-center, i.e. double nuclei that form during the mergers (Fig. \ref{fig:gini_m20}c). \cite{Snyder2015a} show that $G$-$M_{20}$ combination is effective at detecting minor and major mergers for $\sim 0.3$ Gyr in early stages of the merger, while other statistics \citep[e.g., Multipole/Intensity/Deviation, MID;][]{Freeman2013} are more effective at detecting mergers in later stages.

Finally, we note that Gini and $M_{20}$ are sensitive to the method with which the central pixel is calculated. For our control sample, we use the CANDELS morphology catalog computed using the same \textsc{statmorph} code to ensure the datasets are compatible.  

\subsection{Principal Component Analysis}\label{sec:pca}

\begin{figure*}[!htb]
\centering
\includegraphics[width=\textwidth]{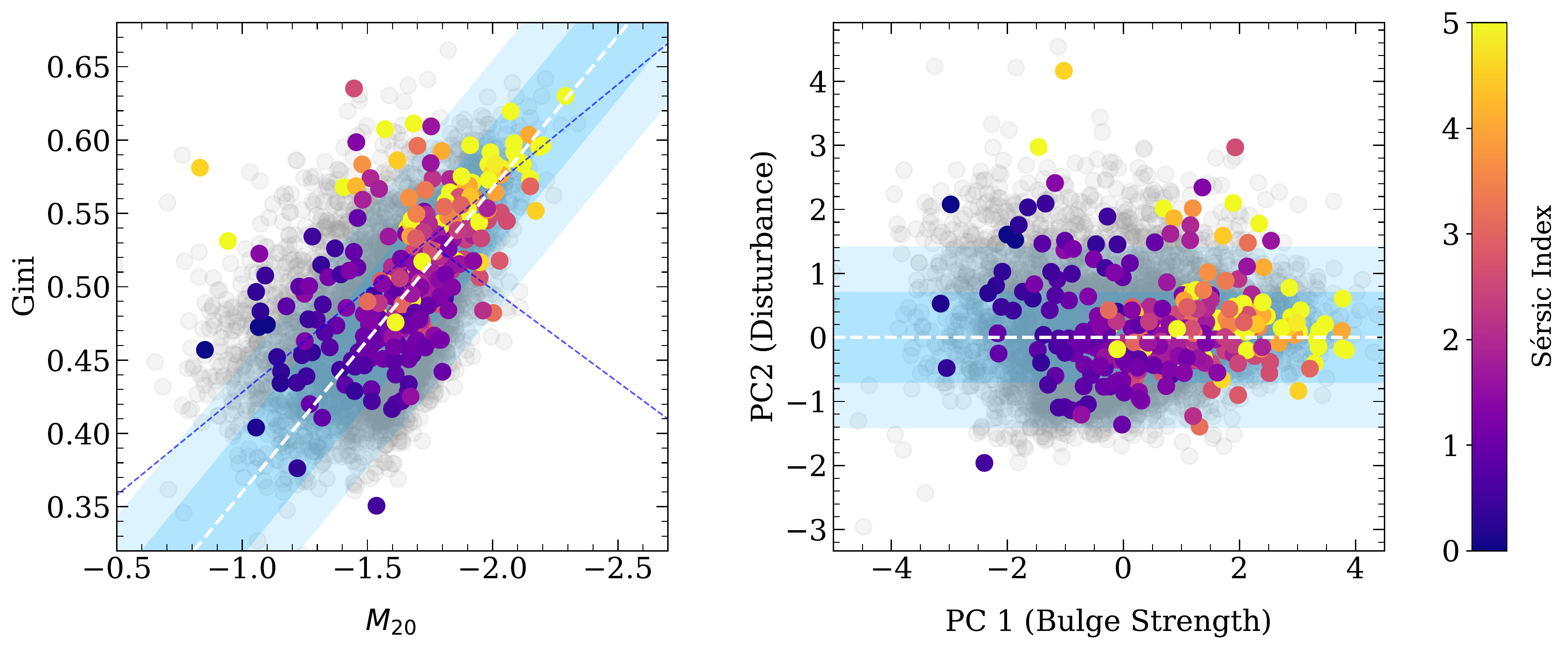}
\caption{\textbf{Left:} Distribution of Gini and $M_{20}$ statistics for the combined sample of photometric members from all 4 clusters. Colored points are the cluster members. The color indicated the galaxy S\'ersic index, ranging from disky (dark blue) to bulge-dominated (yellow). Grey points show all CANDELS field with $1<z<2$ subject to quality cuts in Sec. \ref{sec:qual_cuts}. The white dashed line shows the eigenvector of the first principal component or the Gini-$M_{20}$ main sequence (Eq. \ref{eq:ms}), which clearly traces galaxy bulge strength similarly to S\'ersic index. Shaded regions show $1\sigma$ and $2\sigma$ deviation from the line. Blue dashed lines show the separation of galaxies into early-type (lower left corner), late-type (right region) and disturbed (upper region) galaxies from \citeauthor{Lotz2004} (Eq. \ref{eq:snyder}; \citeyear{Lotz2004}). \textbf{Right:} Principal components of the same distribution. The shaded blue regions show $1\sigma$ and $2\sigma$ deviation in PC2. Again, PC1 is well-correlated with S\'ersic index.}
\label{fig:pca}
\end{figure*}

As discussed above, a combination of $G$ and $M_{20}$ is able to distinguish between disk (low $G$, high $M_{20}$), spheroidal (high $G$, low $M_{20}$) and merging (high $G$, high $M_{20}$) galaxies. \cite{Lotz2004} identify a ``main sequence" on the $G$-$M_{20}$ diagram (shown in blue on Fig. \ref{fig:pca}), a locus of regular galaxies going from disks to spheroids. \cite{Snyder2015b} quantified this using a bulge strength statistic $F(G, M_{20})$, defined as 

\begin{align}
 F(G, M_{20}) = -0.693M_{20} + 4.95G - 3.85
 \label{eq:snyder}
\end{align}

However, this relation was derived for low-redshift galaxies and best matches galaxies with $z<0.4$ \citep{statmorph}.  Therefore, we re-define the bulge strength by locating the Gini-$M_{20}$ ``main sequence" in a CANDELS sample of all $1<z<2$ field galaxies (subject to quality cuts outlined earlier). 

We use Principal Component Analysis (PCA) to find the main sequence. PCA is a statistical tool that uses physical observables to define new variables called principal components (PCs). All PCs are a linear combination of the physical parameters and are orthogonal to each other. Principal components are defined in a descending variance order, so the first PC captures the greatest variance in the data and the last PC captures the least. For a 2-variable parameter space, the variance is maximized along the ``main sequence", so the first PC determines the location on the main sequence and the second PC captures the deviation from that line. Effectively, the first PC always provides the line of best fit in a 2-variable dataset. 

Since $G$ and $M_{20}$ span different scales, they need to be normalized before PCA is performed so that they are weighted equally. For each parameter, we find the mean and the variance of the control distribution. We then subtract the sample mean from each measurement, and divide it by the variance. This produces a normalized dataset.

As shown on Figure \ref{fig:pca}, the first PC (PC1) captures location of a data point along the galaxy ``main sequence". The main sequence is then defined using the PC1 as:

\begin{equation}
    G = -0.21 M_{20} + 0.15
    \label{eq:ms}
\end{equation}

Therefore, galaxies with high Gini and low $M_{20}$ have higher PC1 or bulge strength (Fig. \ref{fig:gini_m20}a), and galaxies with low Gini and high $M_{20}$ have lower PC1  and are more disk-dominated (Fig. \ref{fig:gini_m20}b). As seen on Fig. \ref{fig:pca}, S\'ersic index roughly correlates with PC1: higher values of PC1 correspond to higher S\'ersic indices, bulge-dominated galaxies. Therefore, PC1 successfully captures the bulge strength of a galaxy. Figure \ref{fig:ds9} shows a sequence of galaxies with increasing bulge strength on the right. 

The second PC (PC2) is the perpendicular distance from the galaxy main sequence on a Gini-$M_{20}$ diagram. Galaxies with high PC2 have high Gini and high $M_{20}$ (Fig. \ref{fig:gini_m20}c). As shown in \cite{Lotz2004}, galaxies located anywhere above the Gini-$M_{20}$ main sequence are likely perturbed galaxies undergoing mergers. Therefore, PC2 is a measure of the disturbance of a galaxy. In this work we adapt a stricter definition of a merger candidate, requiring at least PC2 $>$ 1. We note that a negative PC2 does not imply a lower than average disturbance; galaxies with low PC2 correspond to diffuse, poorly resolved galaxies (Fig. \ref{fig:gini_m20}d). Figure \ref{fig:ds9} shows a sequence of galaxies with increasing disturbance on the bottom, and Appendix \ref{app:pca} shows a grid of sample galaxies with various combinations of PC1 and PC2.

\subsection{Quality Cuts}\label{sec:qual_cuts}

We made a number of different cuts to ensure a high quality data sample. We only selected cluster and control galaxies with $m_{H,AB} < 24$, where H-magnitude of the cluster galaxies was calculated by \textsc{SExtractor} using \texttt{MAG\_AUTO} setting.

\textsc{SExtractor} also provides a machine-learning based probability \texttt{p(star)} that a source is a star ($1$) or a galaxy ($0$). All sources with \texttt{p(star)>0.85} were removed unless they were spectroscopically confirmed to be cluster members, in which case they were likely quasars. 

Finally, \textsc{statmorph} provides a flag whenever the morphology parameters are not accurately determined, and all the flagged galaxies were excluded from the analysis. A  galaxy is flagged if the data quality is poor and background level or a Gini segmentation map cannot be properly estimated. Table \ref{tab:clusters} shows the number of flagged galaxies for each cluster. The largest number of flagged galaxies is in the J1432.4 cluster, with 11\% classified as unreliable. In other clusters, fewer than 5\% are flagged. An example of flagged galaxies is shown in Appendix \ref{app:flagged}.

Table \ref{tab:clusters} shows the total number of galaxies left in the analysis after these quality cuts are applied.

\subsection{Statistical Analysis}\label{sec:mcmc}

\subsubsection{Control Sample Selection}\label{sec:control_sample}

The control samples must be constructed in a way such that environmental effects are the only factor determining galaxy morphology. Since morphology is a strong function of redshift, the cluster and the control samples must be redshift-matched. To match the redshift, we selected CANDELS field galaxies within $\Delta z = 0.25$ of the cluster spectroscopic redshift. The number of field galaxies matched the number of cluster galaxies, shown in Table \ref{tab:clusters}.

\begin{figure}
\centering
\includegraphics[width=\linewidth]{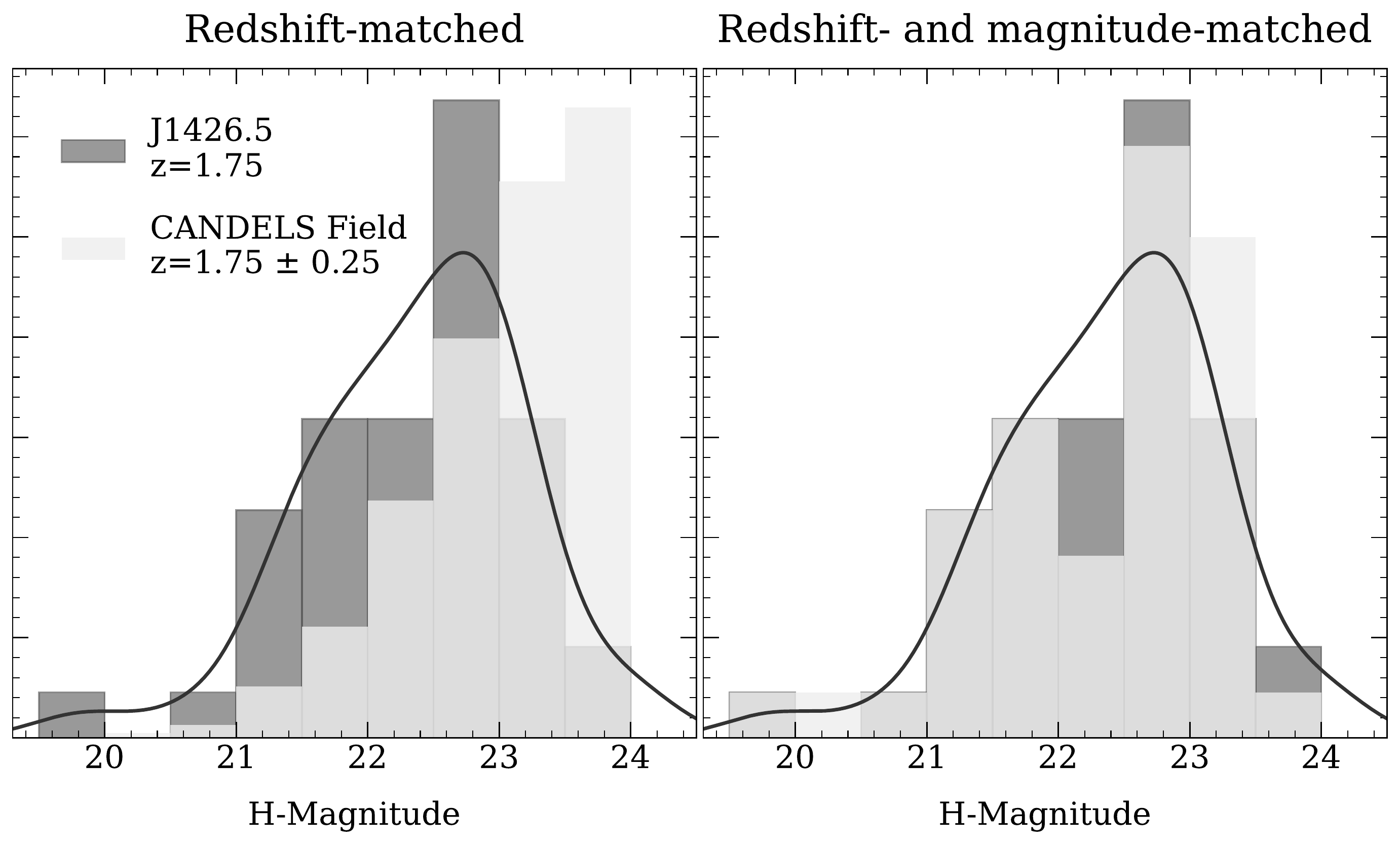}
\caption{An example of one simulation of a magnitude-matched control sample for the J1426.5 cluster (dark grey). The initial redshift-matched sample of field galaxies (left; light grey) has a higher fraction of dim or low-mass galaxies. We fit a Kernel Density Estimate (KDE) to the cluster magnitude distribution, and sample the KDE to select a control sample of the same size as the cluster. The resulting control sample (right; light grey) matches the cluster KDE, therefore eliminating the effects of assembly bias. Initial and final magnitude distributions for all 4 clusters are shown in Appendix \ref{app:mags}.}
\label{fig:idcs_hmag}
\end{figure}

Clusters are also subject to a mass assembly bias \citep[e.g.][]{Mo1996}, where galaxies  in dense environments tend to have larger masses than galaxies in the field. Since massive galaxies quench early on \citep[e.g.][]{Kauffmann2003}, this can bias the comparison between the two environments. As described in Sec. \ref{sec:compactness}, we use H-magnitude as a proxy for stellar mass. 
To correct for assembly bias, we select a control sample that matches the H-magnitude distribution of the cluster galaxies. We estimate the cluster distribution using a Gaussian Kernel Density Estimate (KDE), and then select a control sample that matches the cluster's magnitude KDE. We use KDE in lieu of a binning method to eliminate biases due to selection of bin sizes and edges, and to sample the control galaxies more robustly.

Figure \ref{fig:idcs_hmag} shows an example distribution of magnitudes in the J1426.5 cluster and a control sample at the same redshift. Distributions for all clusters are shown in Appendix \ref{app:mags}. The initial bias towards higher masses in the cluster (dark grey) is evident on the left. We fit the KDE to the cluster distribution (black line), and sample the control sample to match the KDE. The resulting field sample (right, light grey) corrects for the assembly bias.

However, a single randomly selected control sample matching the cluster distribution may not be representative of the entire field population. Therefore, for each cluster we constructed a Monte Carlo (MC) ensemble of 40,000 control samples selected to match the cluster's redshift and magnitude distribution. Each of these samples had the same number of galaxies as the corresponding cluster. We then statistically compared the cluster to the MC ensemble.

\begin{deluxetable*}{rccccccc}[hbt!]
\tablecaption{Fractions of bulge-dominated, disturbed and physically close galaxies in clusters and the control samples}
\tablehead{
\colhead{Cluster Name} & 
\colhead{Average $R_{0.5}$ \tablenotemark{(1)}} &
\multicolumn2c{Bulge Fraction (\%)} &
\multicolumn2c{Disturbance Fraction (\%)} &
\multicolumn2c{Close Neighbors\tablenotemark{(2)} ($\leq 20$ kpc)} \\
\colhead{} & 
\colhead{(kpc)} & 
\colhead{$n>2.5$} & \colhead{PC1$\;>1$} & 
\colhead{$A>0.2$} & \colhead{PC2$\;>1$} & 
\colhead{Number}  & \colhead{Fraction (\%)}} 
\tablecolumns{8}
\startdata
J1142+1527 & $2.0^{+0.3}_{-0.1}$ & 59 & 57 & 11 & 9.7 & 6 & 5.6\\ 
Control & $3.2^{+0.3}_{-0.2}$ & $29^{+5}_{-4}$ & $33^{+5}_{-4}$ & $7^{+3}_{-2}$ & $6^{+3}_{-2}$ & 56 & 2.5 \\ \hline
J1432.3+3253 & $3.4^{+0.1}_{-0.1}$ & 20 & 27 & 30 & 13 & 0 & 0 \\
Control & $3.1^{+0.5}_{-0.4}$ & $30^{+10}_{-7}$ & $30^{+7}_{-10}$ & $7^{+6}_{-4}$ & $7^{+6}_{-4}$ & 92 & 2.3 \\ \hline
J1432.4+3250 & $3.8^{+0.1}_{-0.2}$ & 18 & 27 & 14 & 18 & 28 & 19.6 \\
Control & $3.0^{+0.2}_{-0.2}$ & $24^{+4}_{-3}$ & $24^{+4}_{-4}$ & $8^{+2}_{-3}$ & $8^{+2}_{-3}$ & 60 & 1.6\\ \hline
J1426.5+3508 & $1.9^{+0.15}_{-0.04}$ & 55 & 55 & 4.5 & 4.5 & 4 & 8.9 \\
Control & $3.0^{+0.4}_{-0.3}$ & $34^{+7}_{-6}$ & $30^{+6}_{-7}$ & $9^{+5}_{-4}$ & $9^{+7}_{-3}$ & 35 & 1.2
\enddata
\noindent \tablenotemark{(1)}{Average S\'ersic half-light radius}\\
\tablenotemark{(2)}{Number of galaxies that have a neighbor within a $20$ kpc distance}
\label{tab:fractions}
\end{deluxetable*}

\subsubsection{Monte Carlo Analysis}

For each galaxy from each of the 4 clusters in Table \ref{tab:clusters} we calculated the statistics outlined in Sections \ref{sec:morph} and \ref{sec:pca}. In general, every cluster hosts a wide distribution of morphological parameters. The distribution of each parameter in each cluster is available in Appendix \ref{app:distributions}. 

The aim of this work is to distinguish if the cluster and field populations, rather than individual galaxies, differ, so we compare the median of field and cluster distributions of each morphological parameter.

Since the field sample is large, we can estimate the population median of field galaxies with Monte Carlo sampling. We draw 40,000 samples of field galaxies, compute the median of each sample, and then find the median of the medians -- the population median estimate.

We then check whether the cluster sample is significantly different from the field using a probabilistic approach. We compute the likelihood that the cluster sample could be randomly drawn from the field. If the cluster population is significantly different, then it would be very unlikely to draw a random sample from the field population with a median that is as far, or further, from the population median as the cluster median. 

\begin{figure}[!hb]
\centering
\includegraphics[width=\linewidth]{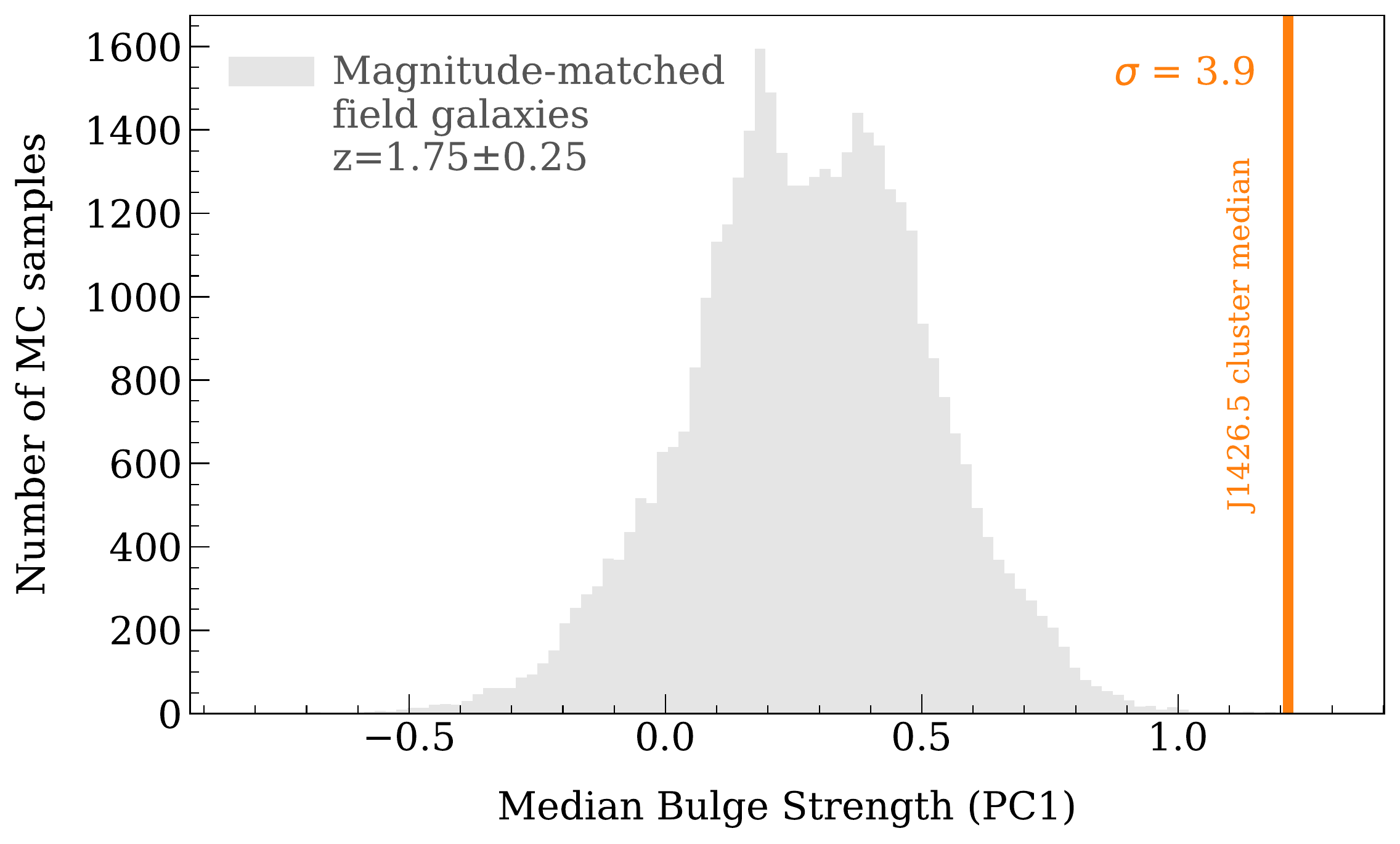}
\caption{The comparison of median bulge strength in J1426.5 cluster galaxies (orange) and 40,000 MC control samples (grey). Only 2/40,000 MC samples have a higher median bulge strength than the cluster, so this result has a $3.9\sigma$ significance corresponding to $p=$2/40,000.} 
\label{fig:mcmc}
\end{figure}

The fraction of field samples satisfying this condition gives the likelihood (a p-value) of randomly drawing the cluster distribution from the field population. We convert the p-value to a significance assuming p-values follow a Gaussian distribution.

Since there is a limited number of MC iterations, the maximum significance is limited by 1 in 40,000 samples with a median equally or more removed from the population median as the cluster, which corresponds to $4.1\sigma$. We chose 40,000 samples, same as in a similar analysis performed by \cite{Hamer2019}, to balance the resolving power of the test and the computational time required for it.

\subsubsection{Galaxy Fractions}

We computed the overall fraction of bulge-dominated and disturbed galaxies for each cluster and the control samples, as well as the number of galaxies that have a close neighbor. 

We defined a galaxy as a ``close neighbor" if it is also a cluster member, and has a projected distance of less than 20 kpc, following \cite{Lotz2013}. \cite{Ellison2010} find that merger fractions are enhanced at separations less than $\approx 40$ kpc in low-redshift field galaxies, making our constraint more conservative. We estimated the corresponding field fraction of the projected ``close neighbors" by selecting all field galaxies subject to quality cuts in Sec. \ref{sec:qual_cuts} and within $\Delta z = 0.25$ of the cluster redshift, and computing the number of galaxies separated by less than 20 kpc in the plane of the sky, with redshift difference less than the photometric redshift uncertainty $\delta z = 0.039$.

We used two different criteria to determine if the galaxy is bulge-dominated: 1) S\'ersic index $n>2.5$, which has been shown to distinguish bulge-dominated and disk-dominated galaxies at high redshifts \citep{Buitrago2013}, and 2) $\textrm{PC1} > 1$. Principal components are defined such that $\textrm{PC} = 0$ is the average PC value for all $1<z<2$ field galaxies, and $PC = 1$ corresponds to 1 standard deviation. So, we labelled all galaxies with $\textrm{PC1} >1$ bulge-dominated as their $\textrm{PC1}$ is at least one standard deviation away from the control sample average. 

We similarly classified galaxy disturbance using Asymmetry and $\textrm{PC2}$. \cite{Peth2016} find using PCA that the most disturbed CANDELS galaxies have $A = 0.2 \pm 0.1$, so we selected disturbed galaxies using 1) $A > 0.2$ or 2) $\textrm{PC2} > 1$.

For the cluster galaxies, the fraction of bulge-dominated or disturbed galaxies is simply defined as a fraction of galaxies of that type to the total number, and therefore has no uncertainty. Since we used 40,000 simulated control samples, we take the median of all 40,000 simulations as the control sample fraction, and assigned the uncertainty using the $16^{th}$ and $84^{th}$ percentiles of the median distribution.

\section{Results}\label{sec:results}

\begin{figure*}[htb!]
\centering
\includegraphics[width=\textwidth]{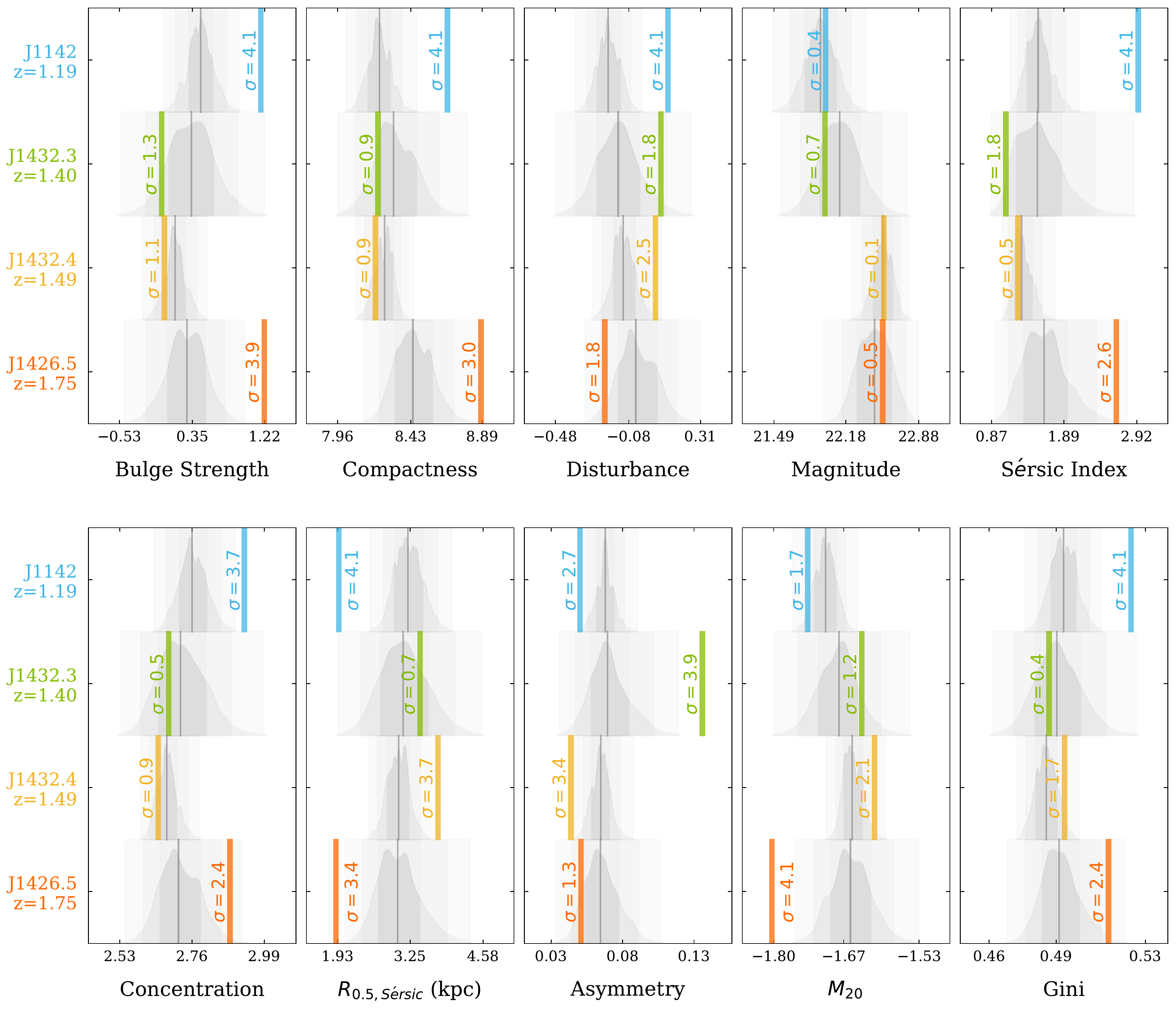}
\caption{The comparison of median morphology parameters between cluster and field galaxies for each cluster in Table \ref{tab:clusters}. For each parameter from Sec. \ref{sec:morph}, the median measurement of cluster galaxies is shown as a colored line. To statistically compare cluster and field galaxies, 40,000 MC simulations of field galaxies matching the cluster redshift and H-magnitude distribution are drawn for each cluster. The distribution of control sample medians is shown in grey, smoothed with a Gaussian KDE. Shading shows $1\sigma$ (dark), $2\sigma$ (medium) and $3\sigma$ (light) intervals of the distribution. Grey solid line shows the population median of field galaxies. Significance of the cluster median is computed as the likelihood of obtaining this median in the MC simulation.}
\label{fig:results}
\end{figure*}

\subsection{Overall Results}

In general, the morphological parameters in all samples have wide distributions, shown in Appendix \ref{app:distributions}. There is a substantial overlap in the distributions of cluster galaxies with their respective control samples, indicating a large variety of morphologies in any environment. However, the aim of this work is to determine whether the morphology distribution statistically differs in cluster and field environments, rather than across individual galaxies.

Table \ref{tab:fractions} shows the fraction of galaxies that are bulge-dominated, disturbed or have a close neighbor, as well as the average galaxy radius for each sample. As seen in the table, J1142 and J1426.5 clusters show a factor of $\approx2$ increase in the number of bulge-dominated galaxies, while J1432.3 and J1432.4 are consistent with the field. On the other hand, J1432.3 and J1432.4 clusters show a significant increase in a fraction of disturbed galaxies.

For a more robust analysis, we compared the morphology of cluster galaxies to field galaxies using medians of each sample, as described in Section \ref{sec:mcmc}. The results are  shown on Figure \ref{fig:results}. 

Each panel shows the distribution of medians of the field sub-samples in grey. The distribution is smoothed by a Gaussian KDE, and shaded regions signify 1$\sigma$ (dark), 2$\sigma$ (medium) and 3$\sigma$ (light) regions. The dark grey line represents the field population median obtained by Monte Carlo sampling. The colored line shows the cluster median for the given parameter. The numerical median values and associated uncertainties for both cluster and field distributions are given in Appendix \ref{app:medians}.

\begin{figure*}[htb!]
\centering
\includegraphics[width=\textwidth]{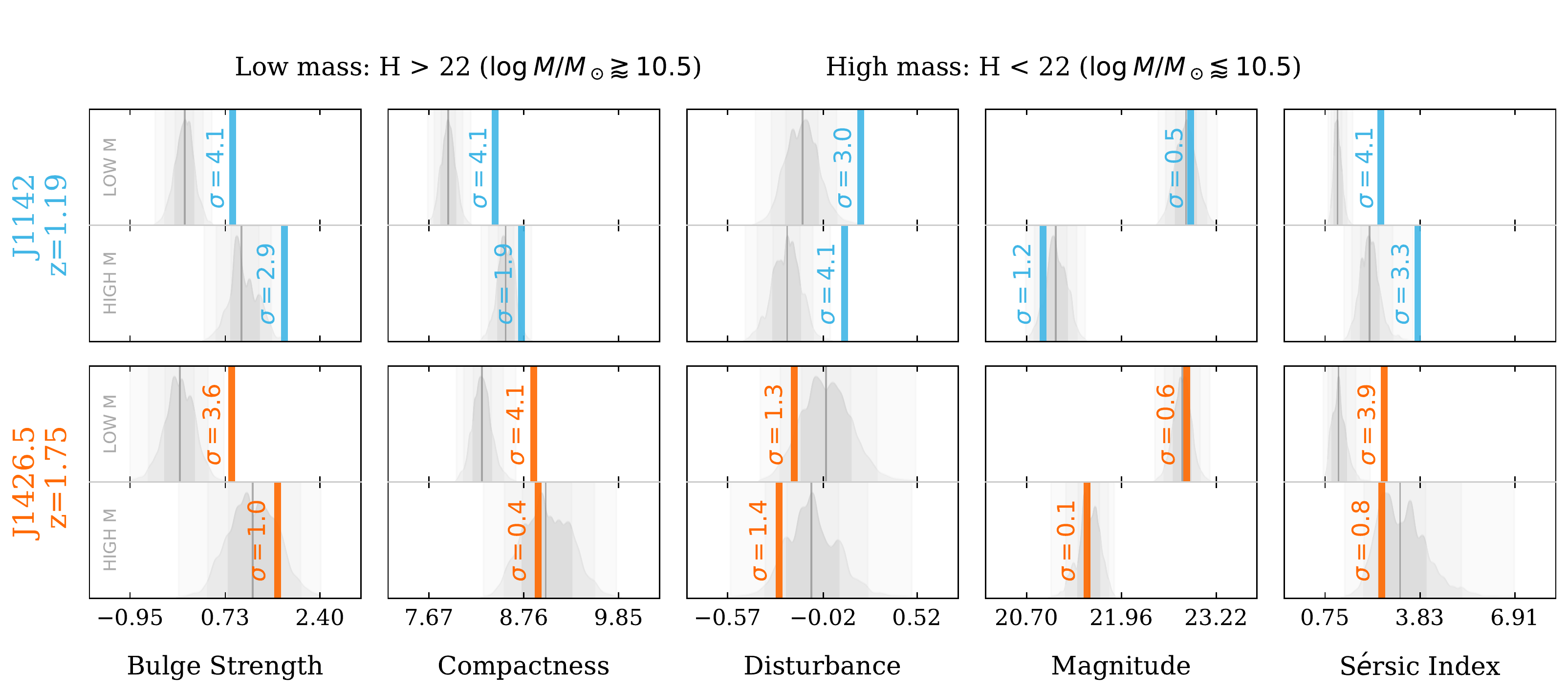}
\caption{Comparison of structural parameters between field (grey) and cluster (colored) galaxies split by H-magnitude as a proxy for stellar mass. Dim and bright galaxies are selected from each cluster and each MC simulation using $H = 22$ as a threshold, which corresponds roughly to $\log M/M_\odot = 10.5$. For each cluster and each morphological parameter, the top panel shows the median distribution of low-mass galaxies, while the bottom panel shows massive galaxies. In both cases, the significance is higher for low-mass galaxies, indicating the stronger influence of cluster environment at a lower mass end.}
\label{fig:mass}
\end{figure*}

Overall, the clusters can be split into two groups: 1) J1426.5 and J1142, that are significantly different from the control sample, and 2) J1432.3 and J1432.4 that are almost indistinguishable from the control sample. First, we will summarize the results for each cluster individually.

\textbf{J1142.} 59\% of galaxies in J1142 are bulge-dominated (Tab. \ref{tab:fractions}), which is 2 times more than in the field. This is also reflected on Fig. \ref{fig:results}. The galaxies in J1142 cluster have larger median bulge strength, compactness and S\'ersic index at the maximum possible significance level, $4.1\sigma$. They also have significantly higher Concentration ($3.7\sigma$) and lower S\'ersic radii ($4.1\sigma$). Gini differs at a $4.1\sigma$ level,  and this is captured in Bulge Strength and Disturbance metrics, which are a linear combination of Gini and M20.  These results indicate that on average, J1142 hosts a much larger population of compact, bulge-dominated galaxies. This is a cluster with a strongly established morphology-density relationship at $z=1.19$. Galaxies in J1142 also have a $4.1\sigma$ higher median disturbance, however this cluster does not have a significantly higher fraction of disturbed galaxies using $A > 0.2$ or $PC2 > 1$ cut-off.

\textbf{J1426.5.} Similarly, J1426.5 shows a factor of $1.8$ increase in the fraction of bulge-dominated galaxies, with 55\% of galaxies showing high values of PC1 and S\'ersic index. Galaxies in this cluster have higher bulge strength at a $3.9\sigma$ level, and are more compact at a $3.0\sigma$ level. Therefore, we confirm that morphology-density relationship exists in some clusters at redshifts as high as $1.75$. We see no significant difference between field galaxies and J1426.5 galaxies in terms of their disturbance or asymmetry.

\textbf{J1432.3.} This cluster, on the other hand, shows no difference from the field in anything but median asymmetry, which is enhanced at a $3.9\sigma$ level. It has an extremely high fraction of asymmetric galaxies, with 30\% of its members having $A > 0.2$. This cluster shows no evidence of a morphology-density relationship being present. It is possible that this is a protocluster, as there is no detection of the hot ICM component in the \textit{Chandra} observation of this region (\textit{Chandra} GO 10457; PI: Stanford), and the increased asymmetry might indicate that the population of bulge-dominated galaxies is being currently built up via mergers. Moreover, the small sample size makes statistical comparison more challenging. This cluster is discussed in more detail in Sec. \ref{sec:iscs}.

\textbf{J1432.4.} Finally, galaxies in J1432.4 are also consistent with the field and show no evidence for the morphology-density relationship. The possible reasons and implications of a lack of the morphology-density relationship in J1432.4 are discussed in Sec. \ref{sec:iscs}. The galaxies in the cluster are more disturbed at a $2.5\sigma$ level, and are less asymmetric at a $3.4\sigma$ level. However, this cluster has a long tail of high-asymmetry galaxies, with 14\% of the cluster galaxies having $A > 0.2$. These enhanced merger signatures indicate that there might be ongoing mergers in this cluster. The low median asymmetry is likely caused by the fact that J1432.4 has shallow imaging of only 2611 s, compared to 8000s exposure time of deep GOODS-S and GOOODS-N CANDELS surveys. Although bulge strength and disturbance, derived from Gini and $M_{20}$, are not very sensitive to imaging depth, Asymmetry is a lot more sensitive \citep{Lotz2004}. 

We now present the dependence of the morphology-density relationship on galaxy mass, morphology and distance from the cluster center. For the rest of our analysis, we focus on the two clusters that exhibit the morphology-density relationship: J1142 and J1426.5, reserving the discussion of the remaining two clusters for Sec. \ref{sec:iscs}. Since bulge strength and disturbance are derived using Gini and $M_{20}$, they are largely insensitive to image depth as long as S/N $>$ 2 \citep{Lotz2004}, so we use these parameters to robustly quantify galaxy morphology. For the remainder of the analysis, we focus on 5 main morphological parameters: bulge strength, compactness, disturbance, magnitude and S\'ersic index. 

\begin{figure*}
\centering
\includegraphics[width=\textwidth]{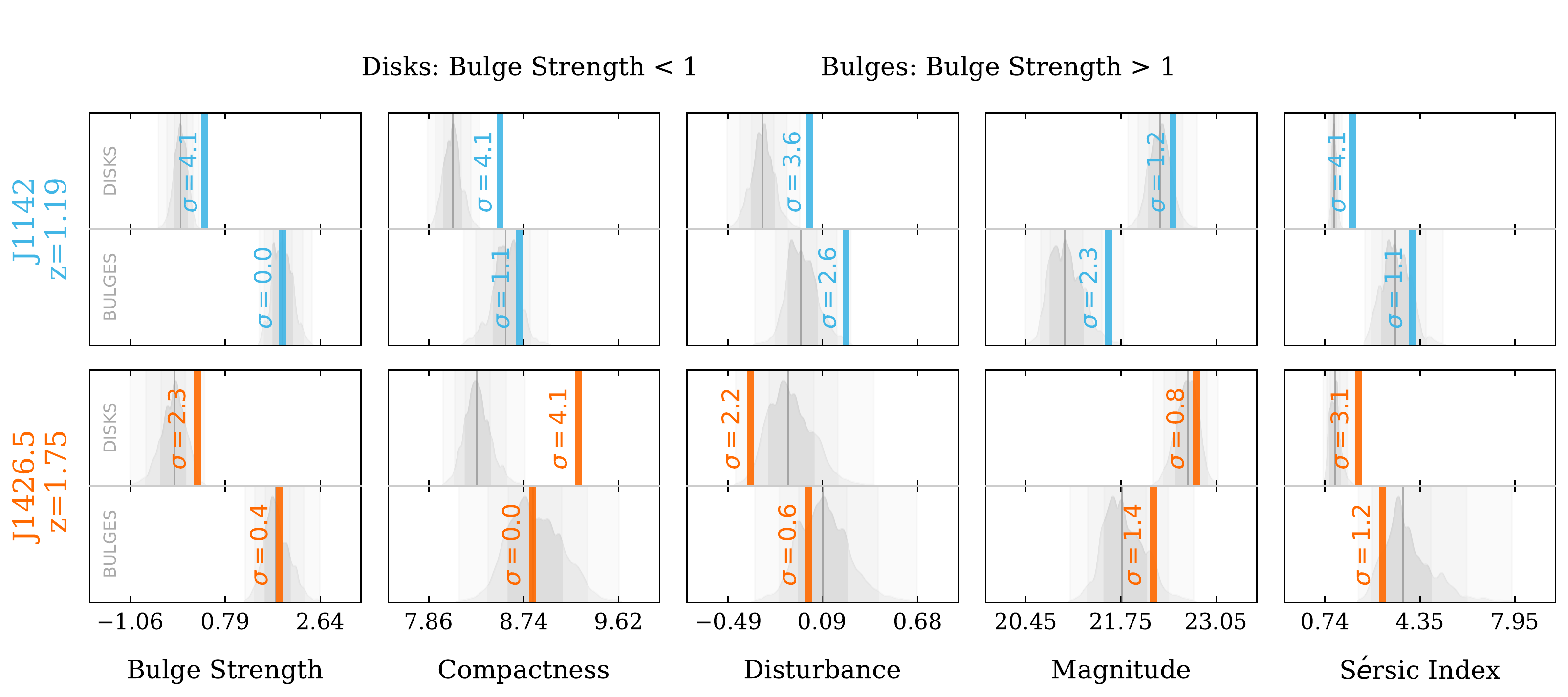}
\caption{Comparison of structural parameters between field (grey) and cluster (colored) galaxies split by galaxy morphology. Bulge-dominated galaxies are selected using Bulge Strength (PC1) $>$ 1 criterion, and remaining galaxies are classified as disk-like. For each cluster and each morphological parameter, the top panel shows the median distribution of disk galaxies, and the bottom panel shows the same for bulge-dominated galaxies. Bulge-dominated cluster galaxies appear consistent with similarly bulge-dominated field galaxies, while disk galaxies have higher median bulge strength and compactness in clusters than in the field.}
\label{fig:morphology}
\end{figure*}

\subsection{Mass Dependence}\label{sec:mass}

To investigate the dependence of morphology on mass, we split the cluster sample into two bins: $H > 22$ and $H < 22$. As seen on Fig. \ref{fig:hmag_mass}, this corresponds roughly to a mass cut-off of $\log_{10} M/M_\odot = 10.5$. Note that at our magnitude cut-off of $H < 24$, the lowest-mass galaxies in the sample have $\log_{10} M/M_\odot \approx 9$. We chose this mass cut as $10^{10.5} M_\odot$ is a common way to separate mass bins in previous studies \citep[e.g.][]{Quadri2012,Cooke2016}. For each MC iteration, we similarly split the control sample into low- and high-mass bins.

Figure \ref{fig:mass} shows the median morphology distributions of the field samples and each cluster. In each panel, the top half shows the distribution of low-mass galaxies ($H > 22$) and the bottom half shows the high-mass galaxies ($H < 22$).

The distributions of low- and high-mass field galaxies are clearly offset from each other. Low mass galaxies have, on average, lower bulge strength, compactness and S\'ersic index. This is to be expected, as massive galaxies are preferentially more red, quiescent  \citep[e.g.][]{vanderWel2014} and hence bulge-dominated in the field, whereas lower-mass galaxies retain their disk morphologies.

In both J1142 and J1426.5, low-mass galaxies are more bulge-dominated and more compact at a $>3.5\sigma$ level. On the other hand, high-mass galaxies are only more bulge-dominated in J1142 at $2.9\sigma$, and not distinguishable from the field in J1426.5. This shows that the cluster environment is more effective at transforming low-mass galaxies compared to the field, whereas morphological evolution of the most massive galaxies is comparable in the field and in clusters.

\begin{figure*}[htb!]
\centering
\includegraphics[width=\textwidth]{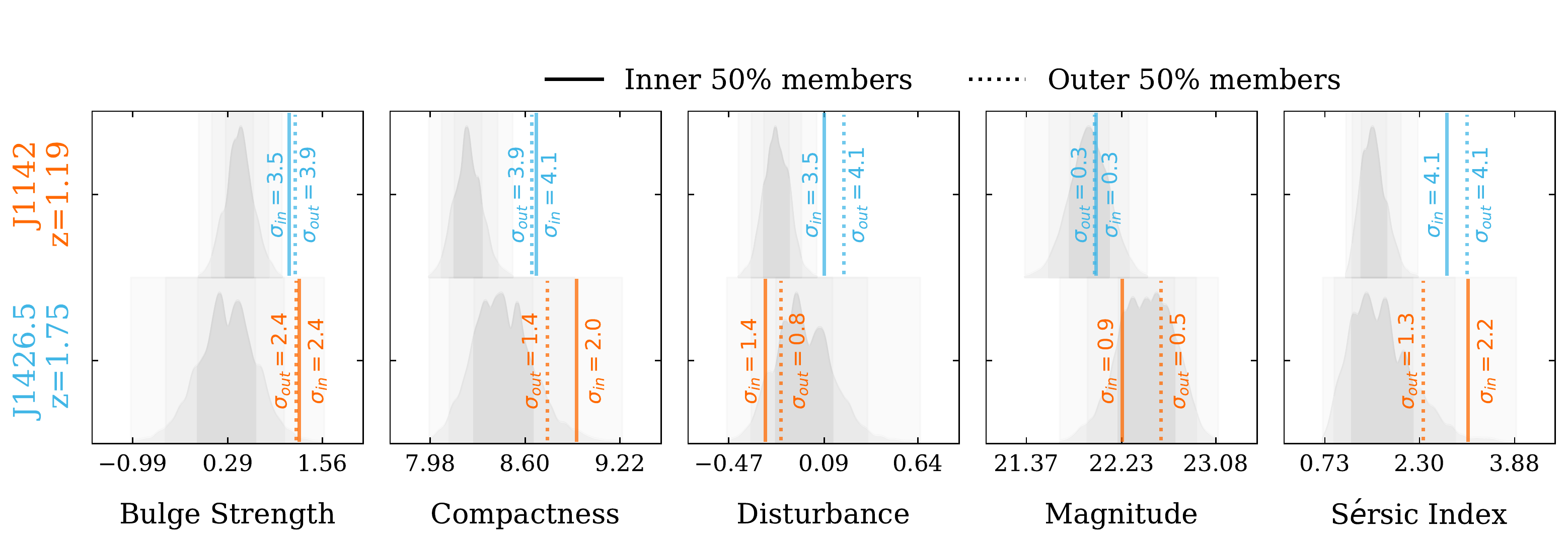}
\caption{Comparison of structural parameters between field (grey) and cluster (colored) galaxies split by clustercentric radius. Cluster galaxies are divided into two groups: inner 50\% (solid colored line) and outer 50\% (dashed colored line). The control sample is similarly split in half randomly. There is no significant difference between the morphology of inner and outer cluster galaxy populations.}
\label{fig:radius}
\end{figure*}

\subsection{Morphology Dependence}\label{sec:morph_dep}

To further investigate how morphology-density relationship is established, we split the galaxies into two groups by their morphology: bulge-dominated (PC1 $>$ 1) and disk-dominated (PC1 $<$ 1). Note that this selection is meant to select out the most bulge-dominated galaxies, while the ``disk" group contains both galaxies with disks and galaxies with intermediate morphologies. We performed this cut for both cluster galaxies and field galaxies in each MC iteration.

Figure \ref{fig:morphology} shows the resulting median distributions of the 5 main parameters. Here, the top panel shows the median distribution of field disks and the bottom panel shows the same for bulges. The corresponding cluster medians are shown in thick colored lines.

Unsurprisingly, the field disk galaxies have lower bulge strength, compactness and Se\'rsic index, indicating that our selection method is effective. Field disks also have higher magnitudes (i.e. are dimmer), showing the effect of mass quenching: massive galaxies tend to have more bulge-dominated morphologies.

There is no significant difference between bulge-dominated cluster and field galaxies for either cluster. Therefore, once a galaxy is bulge-dominated, it does not evolve differently in either field or cluster environments. Considering that we observe a generally larger fraction of bulge-dominated galaxies in clusters compared to the field, we see that galaxy clusters 1) build up a larger population of compact bulge-dominated galaxies via some mechanism, and 2) the compact galaxies produced in clusters are consistent with similarly compact galaxies in the field.

We also find that, although at a lesser significance, bulge-dominated galaxies are on average dimmer in the clusters than in the field. This supports the conclusion of Sec. \ref{sec:mass}, that cluster environment is more important for evolution of low-mass galaxies. Note that because the compactness is the same as in the field but the masses are lower, bulge-dominated galaxies in clusters have smaller radii than in the field, as shown in Table \ref{tab:fractions}.

Finally, we see a significant difference in bulge strength, compactness and S\'ersic index among the disk-dominated galaxies. Disk galaxies in J1142 are  $3.6\sigma$ more disturbed, but not galaxies in J1426.5. In addition, we also find that J1426.5 disk galaxies are  more compact than bulge galaxies. We discuss the likely explanation for these results in Sec. \ref{sec:morph_dep2}.

\subsection{Clustercentric Radius Dependence}

Finally, we investigate the dependence of morphology on the distance from the cluster center. We use the cluster centers as defined in Sec. \ref{sec:data}, and split the galaxies into two groups: innermost 50\% and outermost 50\%. Since clustercentric distance does not matter for field galaxies, we simply split the MC sample in half randomly to construct the control samples for each selection. The resulting distribution of MC medians is the same for both groups, so we use only one of them for the statistical comparison.

Figure \ref{fig:radius} shows the results of this analysis. For each cluster, grey bands show the distribution of the field medians, and the colored lines show the cluster medians. The inner population is represented by a solid line, and the outer population by a dashed one.

As seen on this plot, there is no significant difference between inner and outer population for either of the clusters. Therefore, we conclude that the morphology-density relationship is established evenly throughout the cluster, at least within the $\sim$500 kpc pointing of the HST image, without any morphology gradients.

\section{Discussion}\label{sec:discussion}

\subsection{The Morphology-Density Relationship}

The local Universe displays a strong morphology-density relationship, where the fraction of quiescent, bulge-dominated galaxies increases sharply with local galaxy density, and local galaxy clusters are dominated by early-type galaxies \citep{Dressler1980}. 

Detailed studies of the morphology-density relationship have been limited to lower redshifts due to resolution constraints. \cite{Newman2014} find 90\% significant evidence for morphology-density relationship in a $z=1.80$ cluster, and more recently, \cite{Paulino-Afonso2019} confirm morphology-density relationship in $z < 0.9$ clusters.

To complement morphological studies, SFR is also often used to show the SFR-density relationship. Since SFR calculations rely less on deep optical data, there are more studies into the SFR-density relationship in high-redshift clusters. A number of studies show that this trend continues up to $z < 2$, with cluster galaxies having a factor of $2\sim3$ suppression in star formation efficiency \citep{Quadri2012, Alberts2016, Cooke2016, Lee-Brown2017, Kawinwanichakij2017, Strazzullo2019, vanderBurg2020}. 

 In this work, we found that $\approx50\%$ of J1426.5 and J1142 galaxies are bulge dominated, and the average bulge strength of these galaxies is $4\sigma$ higher than in the CANDELS field.  There is a possibility that some photometric members are interlopers from the field, which would make our sample more consistent with the field than a pure sample, so our significance level might be an underestimation. Therefore, we conclude that the morphology-density relationship is established with a $4\sigma$ significance in some high-redshift clusters, even at redshifts as high as $z=1.75$. 

On the other hand, we found two clusters, J1432.3 and J1432.4, that host a population indistinguishable from the field galaxies. Therefore, we also see that the morphology-density relationship is not evenly established across all clusters at high redshifts. The reasons behind the lack of the relationship in these clusters can be an interesting probe into galaxy evolution, and are discussed in further detail in Section \ref{sec:iscs}. 

Multiple other studies also find a variation from cluster to cluster in terms of the morphology-density and SFR-density relationship \citep{Brodwin2013, Alberts2016}, suggesting that the unique evolutionary pathway of each cluster is more important than the redshift it is found at. In particular, both J1142 and J1426.5 are especially massive clusters for their redshifts, and J1142 is recent cluster-cluster merger \citep{Ruppin2019}. Therefore, it would be extremely useful to repeat this study with a larger sample of 1$<$z$<$2 clusters, and begin to analyze the strength of the morphology-density relationship as a function of cluster properties.

We will now focus on two clusters with a similarly established morphology-density relationship: J1142 and J1426.5. 

We find that the average galaxy size in J1142 and J1426.5 clusters is $\approx2$ kpc, a factor of $0.7$ smaller than the average radius in the field, $\approx3$ kpc. Previous research shows conflicting opinions on the difference between cluster and field galaxy sizes. \cite{Papovich2012} find larger galaxy sizes in a $z=1.62$ cluster with a 90\% significance, and \cite{Delaye2014} find that median sizes in $0.8 < z < 1.5$ cluster galaxies are $10\pm10$\% larger than the field. However, the significance of these results is low. On the other hand, \cite{Newman2014} find no significant difference, and \cite{Matharu2018} find that cluster star-forming and quiescent galaxies are smaller than field galaxies by $0.07\pm0.01$ dex and $0.08 \pm 0.04$ dex, respectively, which agrees with our results. 

This size difference between field and cluster galaxies is further exacerbated by the fact that cluster galaxies are more massive than field galaxies. Due to mass difference alone, galaxies in clusters should be larger, not smaller than in the field. We see this using the compactness metric: cluster galaxies are $4\sigma$ more compact. Compact galaxies are known to quench quickly \citep[e.g.][]{Dekel2013, Zolotov2015}, so the compact galaxies we observe in J1142 and J1426.5 are likely either already quiescent, or progenitors of quiescent galaxies responsible for the SFR-density relationship. 

Since J1142 and J1426.5 are particularly massive clusters, it is interesting to compare their population to clusters of comparable masses at lower redshifts. \cite{Holden2007} find that by z$\approx$0.8 more than 90\% of cluster members are early-type galaxies, while \cite{Smith2005} find that at z$\approx$1, 70$\pm$10\% of cluster members are early-type. The fraction of bulge dominated galaxies in J1142 (J1426.5) is 59\% (55\%), which is consistent with a steady decline of the fraction with redshift.

However, low-redshift studies are based on visual classification of galaxies, and lenticular galaxies are considered early-type. Lenticular galaxies have intermediate bulge strengths \citep{Dressler1980}, so depending on the prominence of their central bulge they may not be included in our bulge-dominated (PC1 $>$ 1) selection. Moreover, the fraction of lenticular galaxies is a strong function of redshift and they may be a missing population in $z>1$ clusters \citep{Dressler1997, Fasano2000}. Considering these two factors, we can conclude that the population of bulge-dominated galaxies is already built up in J1142 and J1426.5 clusters, and the early-type fraction is lower than at lower redshifts due to a lack of lenticular galaxies in our selection. However, this bulge-dominated population is not identical to elliptical galaxies common in local clusters: galaxies in J1142 and J1426.5 are compact, and morphologically more similar to compact spheroids (``red" or ``blue nuggets").

\subsubsection{Dependence on morphology}\label{sec:morph_dep2}

We separated the cluster galaxies into bulge- and disk-dominated using PC1 = 1 as a threshold. We found a significant difference in bulge strength, compactness and S\'ersic index between cluster and field disk galaxies. On the other hand, bulge-dominated galaxies are consistent in the field and in clusters. The fact that the compact galaxies have similar morphologies in clusters and the field suggests that the transformation mechanisms responsible for creating compact bulge-dominated galaxies also act similarly in field and cluster environments. Since there are more compact bulges in clusters, however, this process is more effective there.

On the other hand, we found that disky (PC1 $<$ 1) galaxies in clusters have a significantly different morphology to disky field galaxies: they are more compact, more bulge-dominated and in the case of J1142, more disturbed.  

Since our selection of disk galaxies is less strict than our selection of bulge galaxies, this sample is likely a mixture of true disks and galaxies currently transitioning into a bulge-dominated morphology. Therefore, in cluster environment that are actively transforming their member galaxies, this group is expected to have a larger fraction of galaxies that are currently transitioning towards the bulge-dominated group. This would explain the larger average bulge strength and compactness than in the field. 

The larger disturbance in J1142 suggests that mergers could be responsible for this transition, but the lack of larger disturbance in J1426.5 makes this inconclusive. Alternatively, \cite{Dressler1980} shows that lenticular galaxies have bulge-to-disk ratios that are intermediate between purely bulge-dominated and spiral galaxies, so the intermediate bulge strength population could comprise of lenticular galaxies. However, this is difficult to confirm without color information.

Interestingly, J1426.5 disk galaxies are more compact than the bulge-dominated galaxies. This could be evidence that once the galaxy becomes bulge-dominated, it passively grows via minor mergers and becomes less compact \citep{Naab2009, Wellons2016}, but we see no evidence of this in J1142. 

Alternatively, this could imply that disk galaxies that are inherently more compact survive the passage through the cluster and retain their disk morphology, while extended disks get tidally perturbed and transform into spheroids. This would be expected as the efficiency of tidal interactions depends on the density gradient rather than mass alone \citep{Boselli2006}. However, we do not see evidence of this in J1142. The $2.2\sigma$ offset towards lower disturbance in J1426.5 disks also shows that this effect could be due to a poorer resolution of low-brightness disk galaxies.

Other studies of structural parameters of cluster galaxies separate the galaxies into star-forming and quiescent. Since we expect the bulge-dominated galaxies to quench quickly, we can compare our bulge-dominated sample to the quiescent sample of other studies. \cite{Allen2016} similarly find no difference in either morphology or normalized radius of quiescent galaxies in cluster and field environments. \cite{Matharu2018} find that quiescent cluster galaxies are smaller by 0.08 dex, but since they use physical radii rather than normalized ones, this effect could be a consequence that the bulge-dominated galaxies in our clusters have lower masses than in the field. 

\cite{Socolovsky2019}, on the contrary, find that star-forming galaxies in $0.5 < z < 1$ clusters have larger sizes. However, since our selection of disky galaxies is a likely mixture of disks and intermediate galaxies, and since disk galaxies are not necessarily star-forming in clusters \citep[e.g.][]{Goto2003}, the larger compactness of our disky galaxies does not directly contradict the \cite{Socolovsky2019} result.

\subsubsection{Dependence on mass}

We found that massive ($\log_{10} M/M_\odot > 10.5$) galaxies in the clusters are similar to massive field galaxies, whereas lower-mass galaxies are significantly more bulge-dominated in clusters. This shows that the cluster environment is much more effective at transforming lower-mass galaxies than the field, but not higher-mass ones. 

Other studies of high-redshift make conflicting conclusion on mass-dependence of environmental quenching. \cite{Cooke2016, Lee-Brown2017, Strazzullo2019} find that SFR-density relationship only exists at $1.58 < z < 1.72$ for galaxies with $M \gtrapprox 10^{10.5} M_\odot$, \cite{vanderBurg2020} find that environmental quenching efficiency is 30\% for low-mass galaxies ($10^{9.8} M_\odot$) and 80\% for high-mass ones ($10^{11} M_\odot$), while \cite{Quadri2012} finds no dependence on environmental quenching efficiency on stellar mass. 

On the other hand, in low-redshift clusters, low-mass and satellite galaxies are quenched more efficiently than in the field \citep[e.g.][]{Hogg2003}, so it is not surprising that we also find a higher fraction of compact low-mass galaxies. One reason for the difference between our results and results from high redshift SFR-density studies is that we look at galaxy morphology, rather than star formation. We find that low-mass galaxies have bulge-dominated morphologies, but they might still be star-forming ``blue nuggets" that have not yet been quenched.

In terms of morphology, \cite{Gargiulo2019} find that $0.5 < z < 0.8$ cluster galaxies with $M < 2\times10^{11} M_\odot$ are as compact as field galaxies, while the most massive galaxies are less compact. This would imply that low-mass galaxies evolve similarly in clusters and in the field, while massive galaxies grow in size in clusters, possibly through minor mergers \citep{Naab2009}. We don't see evidence for this growth in our clusters, which might indicate that the period where minor mergers are important starts at $z\approx 1$.

\subsubsection{Dependence on clustercentric radius}

Finally, we see no dependence of the morphology on the clustercentric distance. If morphological transformation was a slow process that happens from the cluster center to its outskirts, we would expect to still see a strong morphological gradient. Therefore, it must either happen early enough in the cluster's history that the gradient is already erased, or galaxies transform upon infall, outside of the HST imaging area.

\subsection{Quenching Mechanisms}

For the remainder of this discussion, we will refer to the mechanism that forms these compact remnants as the ``transformation mechanism" which is closely related to the environmental quenching mechanism, since galaxies that are made compact rapidly quench. In this work, we are primarily focusing on processes that transform the galaxy morphology, and less on those that actually lead to the cessation of star formation.

Our results can be used to constrain the possible mechanism responsible for the morphology-density relationship. The dominant transformation and quenching pathway in galaxy clusters is still disputed, and morphology can be a powerful tool in disentangling the proposed mechanisms, as discussed below. 

\arxiv{\vspace{0.5mm}}

\textbf{Internal Quenching.}  The most massive galaxies quench their star formation early on, regardless of the environment \citep[e.g.][]{Kauffmann2003}, which is also reflected in the steep cut-off at the high-mass end of the Schechter luminosity function \citep{Schechter1976}. Therefore red, quiescent galaxy population is dominated by massive galaxies, while low-mass galaxies are more commonly star forming \citep{Baldry2004}. A variety of internal processes that occur primarily in massive galaxies could lead to this bimodality. \cite{Shankar2004} find that AGN feedback is more effective at masses greater than $10^{10.5} M_\odot$, while \cite{Martig2009} suggest massive galaxies are quenched morphologically: due to deep potential of massive galaxies, the ISM becomes stable against gravitational collapse and stops forming stars. \cite{Martig2009} show that this process is effective at high redshift as well. In this study, we found that the most massive galaxies in the field are more bulge-dominated than the less massive ones, showing again the predominance of massive galaxies in the quenching population. Massive cluster galaxies are more consistent with field galaxies in terms of bulge strength and compactness, especially in J1426.5, therefore we see that the environment plays a lesser role than mass in transforming massive galaxies. On the other hand, we found that lower-mass galaxies are preferentially more bulge-dominated in clusters, showing that the environment is an important factor in galaxy evolution at the intermediate mass range at 1$<$z$<$2.

\arxiv{\vspace{0.5mm}}

\textbf{Ram Pressure Stripping and Strangulation.} In local clusters, environmental quenching is achieved through ram pressure stripping \citep{Gunn1972} or strangulation \citep{Larson1980}. As a galaxy falls into the cluster, the hot ICM removes cold gas from the infalling galaxy and quenches star formation. Since the galaxy's gas is removed and not driven inwards, this process does not build up a compact galactic bulge. Instead, the remnant from ram pressure stripping or strangulation is a red galaxy with a disk or a lenticular morphology. In this analysis, such remnants would appear indistinguishable from the field. However, we found that both J1142 and J1426.5 clusters host more bulge-dominated, compact galaxies that could not have been produced by ram pressure stripping or strangulation alone. Therefore, they cannot be the only quenching mechanisms in high-redshift clusters.

Similarly, \cite{Dressler1997} and more recently \cite{Cerulo2017} found that $z\approx0.5$ clusters have a larger fraction of elliptical, but not lenticular galaxies than the field. They concluded that elliptical galaxies are built up early during cluster formation, whereas S0s appear only after the cluster is virialized. \cite{Balogh2016} propose that ram pressure stripping is not a dominant quenching mechanism in $0.8 < z < 1.2$ clusters. This agrees with our conclusion that at 1$<$z$<$2, compact population is built up by a different mechanism, which becomes dominated by ram pressure stripping and strangulation at lower redshifts.

\arxiv{\vspace{0.5mm}}

\textbf{Stripping of Compact Galaxies.} A similar environmental quenching mechanism has been proposed by \cite{Socolovsky2019}\, based on a ‘bathtub’-type model \citep{Bouche2010}. In a normal galaxy, feedback from star formation and AGN blows the gas out from the galactic disk into the circumgalactic medium (CGM), and this gas is then recycled back into the disk. As the galaxy enters a cluster, ram pressure from the ICM can further trigger the AGN, increasing the amount of gas fed into the CGM \citep{Poggianti2017}. The hot ICM easily strips the loosely bound gas from CGM, therefore inhibiting recycling. Since compact galaxies have higher star formation surface density, they expel a larger fraction of their ISM in outflows \citep{Heckman1990}, so more of their gas supply is in the CGM which is easily stripped by the ICM. Therefore, in-situ compact galaxies would be preferentially quenched in clusters. If this is the dominant quenching pathway, we would expect to find that 1) quiescent galaxies are more compact, 2) star-forming galaxies are more extended and 3) the overall fraction of compact galaxies is the same as in the field. \cite{Socolovsky2019} confirms 1) and 2) by showing that red cluster galaxies at $0.5<z<1$ are preferentially more compact than their star-forming counterparts. However, our analysis does not consider the star formation state of the galaxy, solely the morphology, and we see an overall higher fraction of compact galaxies in J1142 and J1426. The formation of compact galaxies in clusters cannot be explained by this quenching mechanism alone, and a different environment-dependent mechanism must be in place that actually makes infalling galaxies more compact.

\arxiv{\vspace{0.5mm}}

\textbf{Violent Disk Instability.} To create a compact galaxy, a physical process must create an instability in the galactic disk to drive the gas inwards and build up the galactic bulge. A violent disk instability (VDI) is one such proposed process \citep{Dekel2013, Zolotov2015}. VDI is triggered by a high flux of cold gas from the cosmic web into the galactic disk, causing it to become unstable to perturbations and collapsing. This would produce a prominent galactic bulge, and lead to quenching when the central gas is expelled via AGN and/or star formation feedback. This mechanism would be effective in the field, but is likely suppressed in galaxy clusters with a hot ICM. Both J1142 and J1426 have significant X-ray emission from the ICM, so VDI is unlikely to be effective at the redshifts we observed. However, the supply of cold gas to clusters could have been enhanced earlier during the cluster formation, before it is virialized. VDI could have been a viable quenching mechanism at higher redshifts and away from the cluster center. Therefore, it might still be responsible for building up the population of compact galaxies we found, and even higher redshifts or protoclusters must be studied to constrain the effect of VDI.

\arxiv{\vspace{0.5mm}}

\textbf{Mergers.} Another mechanism often invoked to produce compact galaxies is a major wet merger \citep{Hopkins2009, Wuyts2010, Wellons2015}. It causes the gas of the merging galaxies to collide, lose its angular momentum and stream to the galactic center. Therefore, the bulge is built up, and eventually bulge gas is expelled. Due to a high velocity dispersion ($500-1000$ km/s) of cluster members, mergers are very uncommon in local clusters. However, \cite{Lotz2013} and \cite{Watson2019} find a higher merger rate in high-redshift clusters: a 57\% rate at $z=1.62$ (compared to 11\% in the field) and a 10\% rate at $z=2$ (compared to 5\% in the field).

If major mergers are the dominant transformation mechanism in high-redshift clusters, we would expect to see 1) a large population of compact bulge-dominated galaxies, and 2) an enhancement in merger signatures in cluster galaxies. We do find a factor of $2$ increase in the number of bulge-dominated galaxies compared to the field, and an increased disturbance in the J1142 cluster, but no increase in the number of projected close neighbors. J1426.5 shows the same number of merger candidates as the field.

However, a lack of strong merger signatures in J1142 and J1426.5 does not rule out mergers as a viable transformation mechanism. Since we already found a significant increase in the number of bulge-dominated galaxies, the majority of mergers that could be responsible have already happened. Remaining merger signatures might no longer be detectable in high-redshift clusters. \cite{Lotz2004} shows that disturbance is detectable even in images with low SNR, but \cite{Snyder2015a} shows that this traces intermediate, rather than late, stage mergers. Late stage mergers of star-forming galaxies are detected by low surface brightness features such as tidal tails,  well traced by Asymmetry \citep{Snyder2019}, which is much more reliant on deep imaging and high SNR \citep{Lotz2004}. Therefore, to constrain the importance of mergers, we require deeper imaging of established clusters,  imaging of high-redshift clusters and protoclusters where mergers might still be taking place, and imaging of cluster outskirts where infalling galaxies might be actively merging. In fact, we find an increase in merger signatures in J1432.3 and J1432.4 that do not have an established morphology-density relationship yet, supporting that an earlier epoch of galaxy mergers might be responsible.

A challenge to the merger/VDI mechanism is that we found that low-mass galaxies are transformed preferentially, and neither VDIs not mergers would necessarily cause this effect. However, it is possible to propose a mechanism involving a merger/VDI where this is the case. Simulations show that some of the merging gas builds up the bulge, and some can reform the disk \citep[e.g.][]{Hopkins2009b}. The disk would not reform if the bulge gas is stripped away by the ICM ram pressure. It is easier to strip a low mass galaxy due to its shallow potential, so mergers or VDIs in tandem with ram pressure stripping could lead to a higher fraction of bulge-dominated low-mass galaxies that we observed.


\arxiv{\vspace{0.5mm}}

\textbf{Tidal Interactions.} Tidal interactions between galaxies during fly-bys can also drive the gas inwards, building up a bulge component. However, simulations of local clusters show that these fly-bys are too rapid, and ineffective in disturbing the star-forming disk \cite{Byrd1990, Fujita1998, Boselli2006}. Although high-redshift clusters have lower velocity dispersion implying longer interaction timescales, \cite{Villalobos2014} shows the individual encounters are inefficient even in galaxy groups, so it is unlikely that they are important in clusters at any redshift. However, multiple rapid encounters known as harassment can contribute to the galaxy transformation in clusters \citep{Moore1996}. 

On the other hand, interaction of a galaxy with the cluster potential could be sufficient to drive the gas radially into the bulge \cite{Byrd1990, Valluri1993}. The gas can be tidally driven inwards for a sufficiently large perturbation parameter, defined as

\begin{equation*}
    P_{gc} = (M_{c}/M_{g}) (r_{g}/R)^3
\end{equation*}

where $M_{c}$ and $M_g$ are masses of the cluster and a galaxy, $r_g$ is the radius of the galaxy and $R$ is the distance of that galaxy to the cluster center. \cite{Byrd1990} show that such tidal perturbations are effective for $P_{gc} \geq 0.006-0.1$. \cite{Boselli2006} provide a simple estimate for the Coma cluster, and show that the cluster potential is strong enough to cause inflows in the infalling galaxies. 

We can repeat this estimate for J1142 and J1426.5 clusters. We parameterize the cluster mass as $M_c = M_{14}^c \times 10^{14} M_\odot$, the galaxy mass as $M_g = M_{10}^g \times 10^{10} M_\odot$, where $M_{14}^c = 4\sim11$ and $M_{10}^g=0.1\sim10$. We use the average radius of a field galaxy, $3$ kpc. For the clustercentric distance, we use the typical size of the HST image, $500$ kpc, since the morphology-density relationship appears to exist everywhere in the imaged region. Then the interaction efficiency is

\begin{equation*}
    P_{gc} \approx (M_{14}^c/M_{10}^g) \times 0.002
\end{equation*}

Plugging in $M_{14}^c=4$ for the J1426.5 cluster, we see that $P_{gc}$ only exceeds $0.006$ for galaxies with $M_{10} < 1.1$. On the other hand for J1142 cluster with $M_{14}^c=11$, $P_{gc}$ exceeds $0.006$ for $M_{10} < 4$. Therefore, we would expect that only lower-mass galaxies are tidally perturbed in J1426.5, whereas low- and high-mass galaxies are affected in J1142. This is indeed what we found in Sec. \ref{sec:mass}.

Tidal interaction with the cluster potential is an attractive candidate for the environmental quenching mechanism for multiple reasons. 1) It is capable of forming our observed compact, bulge-dominated galaxies. 2) It is more effective for low-mass galaxies, since massive galaxies are stabilized by their self-gravity. This explains our finding that low-mass galaxies are transformed preferentially in clusters. 3) Since this mechanism is more effective at perturbing more extended objects, it would explain our result that surviving disks in J1426.5 are extremely compact. And importantly, 4) this mechanism is unique to cluster environments. Unlike with other mechanisms, such as VDI and mergers, we don't need to invoke an additional reason as to why it's more effective in clusters than in the field.

However, tidal interactions would likely result in a strong gradient of the morphology-density relationship, since the cluster potential is deeper in the cluster center, and we do not observe such a gradient. Moreover, the interaction efficiency estimate of $0.006$ was derived from older studies, when the entire cluster potential was not simulated at all to allow better resolution for the galactic disk. Although tidal interactions are shown to be important, careful modelling is required to determine whether they lead to radial inflows and the build-up of galactic bulges. More recently, \cite{Villalobos2012} showed that this is not the case in collisionless N-body simulations of groups. To confirm whether tidal interactions are be important, it is crucial to repeat the efficiency a measurement with new, state-of-the-art hydrodynamical simulations of galaxy clusters, such as Illustris \citep{Vogelsberger2014}, EAGLE \citep{Schaye2014,Crain2015}, The Three Hundred Project \citep{Cui2018} and others.

\arxiv{\vspace{0.5mm}}

\textbf{Pre-processing.} Group pre-processing is an emerging, but important mechanism that could cause a morphology-density relationship in clusters. Cosmological simulations show that at $z\approx2$ clusters consist of a $\sim$10 Mpc wide network of smaller galaxy groups, that will merge into the cluster halo by $z\approx0$ \citep{Muldrew2015}. Therefore, \cite{Muldrew2015} and \cite{Hatch2016} argue that the evolution of an infalling galaxy in the group stage is important, and significant pre-processing may occur there. \cite{Dressler2013} also find that a substantial number of galaxies infalling in the 0.3$<$z$<$0.5 clusters is being pre-processed in groups. Galaxy groups are smaller, less massive and have lower velocity dispersions. Therefore, they are less hostile to VDI and galaxy mergers, while tidal interactions could still be important. 

We see that the morphology-density relationship is already established evenly within the 500 kpc imaged region of the J1426.5 and J1142 clusters. Since the clusters are young, if the relationship established inside-out (i.e. via tidal forces), it would have to happen very rapidly. On the other hand, group pre-processing suggests that the bulge-dominated galaxies are already formed in groups when the galaxies infall. This would produce a flat morphology-density gradient on the scales that we observe. To further investigate the effect of group pre-processing in the cluster outskirts, wider-field imaging of the cluster is necessary.

\subsection{J1432.3 and J1432.4 clusters}\label{sec:iscs}

Another interesting probe into the origin of the morphology-density relationship is the J1432.3 and J1432.4 pair that hosts galaxies consistent with the field. 

J1432.3 is not detected in X-ray, although it has been observed by \textit{Chandra} GO 10457 (PI: Stanford) in the same pointing as J1432.4. Moreover, out of 30 members detected for J1432.3, only 8 are within 500 kpc of the cluster center. Such a low galaxy number density, a lack of both a substantial hot gas reservoir and a morphology-density relationship all suggest that this is still a protocluster or a group. 

As discussed above, studying groups and proto-clusters is essential to further constrain the mechanism that forms compact galaxies in clusters. J1432.4 shows a $4.1\sigma$ enhancement in average asymmetry, and 30\% of the cluster galaxies have $A>0.2$ compared to $7^{+6}_{-4}\%$ in the field, indicating a much higher merger rate than in the field. Therefore, this cluster suggests that mergers are important in such young dense environments.

On the other hand, J1432.4 is a cluster with a detected ICM component, a mass of $(2.5 \pm 0.5) \times 10^{14} M_\odot$ and an established red-sequence. Judging by this, J1432.4 appears to be in a similar evolutionary stage to J1142 and J1426.5. I is surprising that this cluster shows no evidence of the morphology-density relationship. If this difference is indeed physical, rather than an observational effect, there must be a factor other than cluster mass or a presence of hot ICM that determines galaxy evolution within the cluster.

Determining whether a group of galaxies forms a cluster or a protocluster is difficult, and the lack of a morphology-density relationship can suggest that J1432.4 might still be in the process of assembly. Since the cluster shows high mass, red sequence and X-ray emitting ICM but no morphology-density relationship, the timescales that govern the morphology-density relationship and other metrics of cluster establishment must differ. If that is the case, morphology can be another powerful probe in quantifying the evolutionary stage of a cluster.

However, this cluster has the most shallow exposure of $2,611$s per pointing out of all four clusters, and the cluster membership is determined via red sequence selection rather than photometric redshifts. This adds significant uncertainty in our analysis of J1432.4 galaxies. First, the imaging might not be sufficiently deep to resolve concentrated bulge light and diffuse disks. Secondly, red sequence selection can introduce contamination from the field, therefore reducing the signal of the morphology-density relationship.

Finally, these two clusters exist in a rare context. J1432.4 is only separated from J1432.3 by 3$'$ (1.5 Mpc) on the sky, and by $\approx73$ Mpc in 3D space. High-redshift clusters are better described as extended network of groups, often sprawling on a $30\sim40$ Mpc scale \citep{Muldrew2015}, which will then infall into the main cluster halo. Therefore, there is a significant chance that the group populations of J1432.4 and J1432.3 overlap. Moreover, at high redshift, forming clusters are embedded in a complex web of filaments \citep{Cautun2014}, so a filament could connect these two clusters. Lastly, due to innate uncertainties in photometric and red-sequence membership estimation, the sample population of J1432.4 is certainly contaminated by J1432.3 and filamentary galaxies.

All these factors make J1432.4 and J1432.3 an incredibly complex system to study and could explain why J1432.4 appears to be inconsistent with J1142 and J1426. Contamination from the cosmic web galaxies and J1432.3 could erase the signal from the morphology-density relationship, so a thorough statistical decontamination is required to study this system. Moreover, the imaging depth of J1432.4 is significantly shallower compared to other clusters, and morphology estimates might be insufficiently accurate. Therefore, additional imaging of this region is necessary to make conclusions about these clusters.

Finally, and perhaps most curiously, the interaction between the two clusters could potentially lead them down a very different evolutionary path. If a filament connects the two clusters, the gas inflow along the filament may fuel additional star formation in these clusters and maintain disk-dominated, star-forming galaxies. \cite{Chen2017} show that galaxies along cosmic filaments have larger sizes than field galaxies. We similarly find that J1432.3 has an average galaxy size of $3.4^{+0.2}_{-0.1}$ kpc, slightly larger than $3.1\pm0.4$ kpc in the field, and J1432.4 galaxies have an average size $3.8^{+0.1}_{-0.1}$ kpc, larger than the field average of $3.0^{+0.4}_{-0.3}$ at a $3.6\sigma$ level. Therefore, studying the evolution of this cluster independently of J1432.3 is impossible, and they need to be carefully considered together instead. A deep study of these two clusters and the region between them is necessary to shed light on structure formation as well as morphology-density relationship in this unique setting. 

\section{Conclusion}\label{sec:conclusions}

We studied the morphology of galaxies in 4 $1<z<2$ galaxy clusters using deep HST imaging and open-source morphology code  \textsc{statmorph}. 

First, we established that a morphology-density relationship exists as far back as $z=1.75$. Two out of four galaxy clusters studied have a 50\% fraction of bulge-dominated galaxies, and an enhanced average bulge strength at a more than $3\sigma$ significance level. This means that the population of spheroidal galaxies is built up very early in the Universe in sufficiently dense environments.

We also found that in these two clusters, there are more compact galaxies than in the field. Therefore, there must be some environment-dependent mechanism that not only quenches galaxies, but also transforms them into compact, bulge-dominated remnants. It is unlikely to be the same environmental quenching mechanism as in the local Universe, such as ram pressure stripping or strangulation, since they would not result in the formation of compact, bulge-dominated galaxies.

Finally, we saw that it is the lower-mass ($\log_{10} M/M_\odot < 10.5$) galaxies that are preferentially quenched in clusters, especially in J1426.5. This shows that on the higher mass end, mass quenching is still the dominant process and is independent of the environment; whereas on the low mass end, environmental quenching is important.

We found no evidence of a gradient in the morphology-density relationship, at least within the $\sim$500 kpc HST region that we studied. Therefore, the quenching mechanism is unlikely to act from the inside-out on a smaller scale than 500 kpc.

All these factors point to an environmental quenching mechanism that 1) is enhanced in dense environments, 2) transforms a galaxy's morphology to a compact bulge, 3) affects preferentially low mass galaxies, and 4) is effective out to at least 500 kpc from the cluster center.

Tidal interactions with the cluster's potential fit requirements 1)-3). VDIs and wet mergers could fit these requirements as well, but assumptions need to be made to explain an enhanced rate of these events in clusters and the preferential quenching of low-mass galaxies. Finally, group pre-processing fits all of the above requirements, especially considering that the majority of cluster galaxies goes through a group stage at some point \citep{Muldrew2015}.

In addition, we found a rare z$\approx$1.45 cluster \textit{complex} formed by J1432.3+3253 and J1432.4+3250 clusters that are separated by $\approx$73 Mpc. These two clusters show no evidence of the morphology-density relationship, but J1432.4 shows an increased asymmetry, implying a high merger rate, and possibly the beginning of morphological transformation of the cluster's galaxies. 

Since J1432.4 is a relatively massive cluster with a hot ICM and a red sequence, the lack of the morphology-density relationship in J1432.4 could indicate that this relationship depends more than on the cluster mass or ICM, and shed light on the physical process responsible. If that is the case, galaxy morphology could act as another metric of cluster establishment in addition to cluster mass, and become a useful tool in proving the evolutionary state of a cluster or a galaxy cluster.

However, J1432.3 and J1432.4 form a complex cluster system, where cluster membership could be contaminated both by the other cluster and any filaments connecting them. Evolution of galaxies in the filaments could also proceed differently from groups or clusters. A further study of this complex will most certainly shed light on questions regarding structure formation and galaxy evolution in the most extreme environments in the Universe.

\section*{Acknowledgements}

We thank the anonymous referee for the extremely useful comments regarding this manuscript. We also thank Roan Haggar for fruitful scientific discussions, and Jacob Hamer for helpful comments on the statistical basis of this work. This work is based on observations made with the NASA/ESA Hubble Space Telescope, and obtained from the Hubble Legacy Archive, which is a collaboration between the Space Telescope Science Institute (STScI/NASA), the Space Telescope European Coordinating Facility (ST-ECF/ESA) and the Canadian Astronomy Data Centre (CADC/NRC/CSA). In addition, much of this work is based on observations taken by the CANDELS Multi-Cycle Treasury Program with the NASA/ESA HST, which is operated by the Association of Universities for Research in Astronomy, Inc., under NASA contract NAS5-26555. This research made use of Astropy, a community-developed core Python package for Astronomy \citep{astropy}.

\bibliographystyle{mnras_custom}
\raggedright

\vspace{-4mm}
\setlength{\bibsep}{0mm}
\bibliography{main}

\appendix

\section{Sample galaxies spanning PCA space}\label{app:pca}

We use PCA to classify galaxy morphology, as shown in Sec. \ref{sec:pca}. The first principal component, PC1, represents the bulge strength of the galaxy and correlates with S\'ersic index. The second principal component, PC2, is shown to indicate a recent merger if high \citep{Lotz2004}. A low PC2 indicates a diffuse morphology. The mosaic below shows HST images of some galaxies from either one of the 4 clusters in Tab. \ref{tab:clusters}. We used a square root scaling to highlight low-brightness features, such as merger signatures. Note that because of the scaling, the bright cores of high-PC1 galaxies are de-emphasized. 

\begin{figure*}[h!]
\centering
\includegraphics[width=0.9\textwidth]{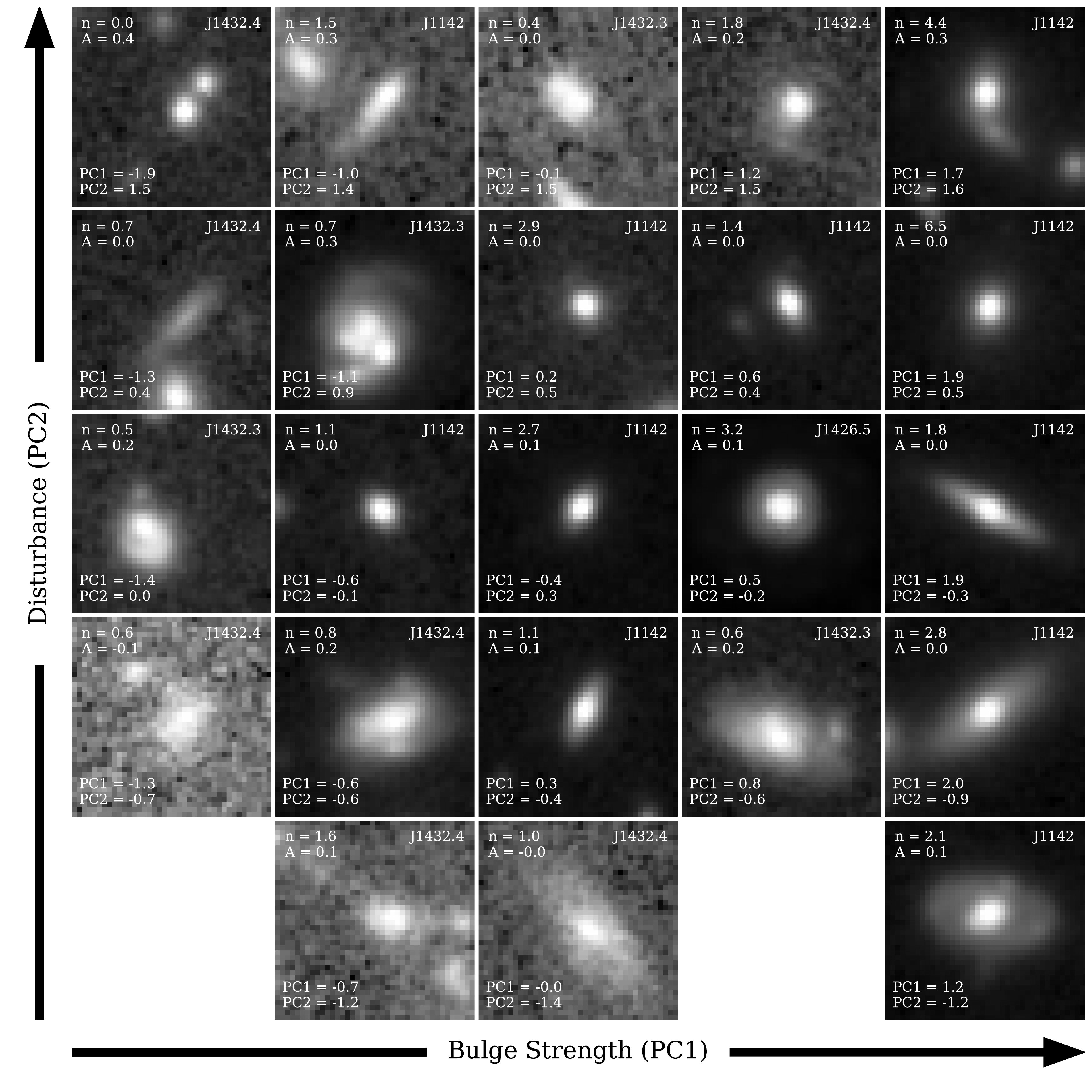}
\caption{Example Distribution of the morphological statistics outlined in Sec. \ref{sec:morph} for each cluster in Table \ref{tab:clusters}. Three distributions are shown: spectroscopic cluster members (dark red), photometric members (light red) and the control sample (grey). Dotted grey lines indicate the distribution median for cluster members (red) and the control sample (grey).}
\label{fig:pca_space}
\end{figure*}

\section{Galaxies flagged by \textsc{statmorph}}\label{app:flagged}

\textsc{statmorph} provides a quality flag for galaxies where the computed morphologies are unreliable. Generally, few galaxies are flagged, and the total number of flagged galaxies for each cluster is summarized in Table \ref{tab:clusters}. The primary reason a galaxy would be flagged is if its surface brightness is too low, and therefore the random noise from the background dominates the galaxy light, causing the segmentation map to be irregular. An image can  also be flagged if it contains two nearby but not overlapping objects, while the segmentation map falsely identifies them as a single galaxy.

\begin{figure*}[h!]
\centering
\includegraphics[width=\textwidth]{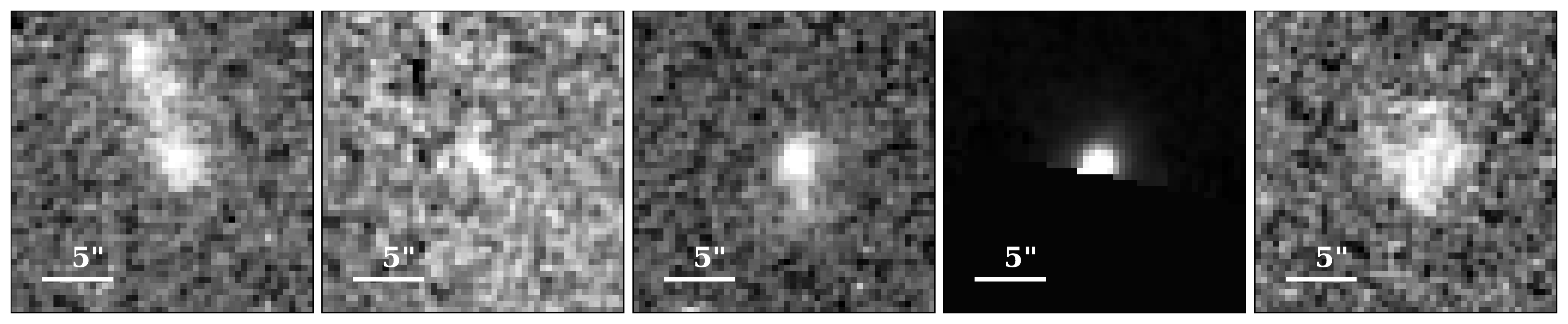}
\caption{A sample of galaxies flagged by \textsc{statmorph} as having unreliable morphology measurements. Most galaxies are too dim, and therefore their segmentation maps are dominated by the random background fluctuations. The fourth galaxy is bright, but is located at the edge of the \textit{HST} pointing, and therefore the galaxy image is cut off.}
\label{fig:flagged}
\end{figure*}

Figure \ref{fig:flagged} shows a sample of flagged galaxies from this study.

In addition, \textsc{statmorph} provides a S\'ersic flag when non-parametric morphologies were computed successfully, but a S\'ersic fit failed. This can happen in disturbed galaxies, when the stellar light is not well-described by a 1-component Se\'rsic fit.

\section{Magnitude-Matched Control Samples}\label{app:mags}

\begin{figure*}[b!]
\centering
\includegraphics[width=0.8\textwidth]{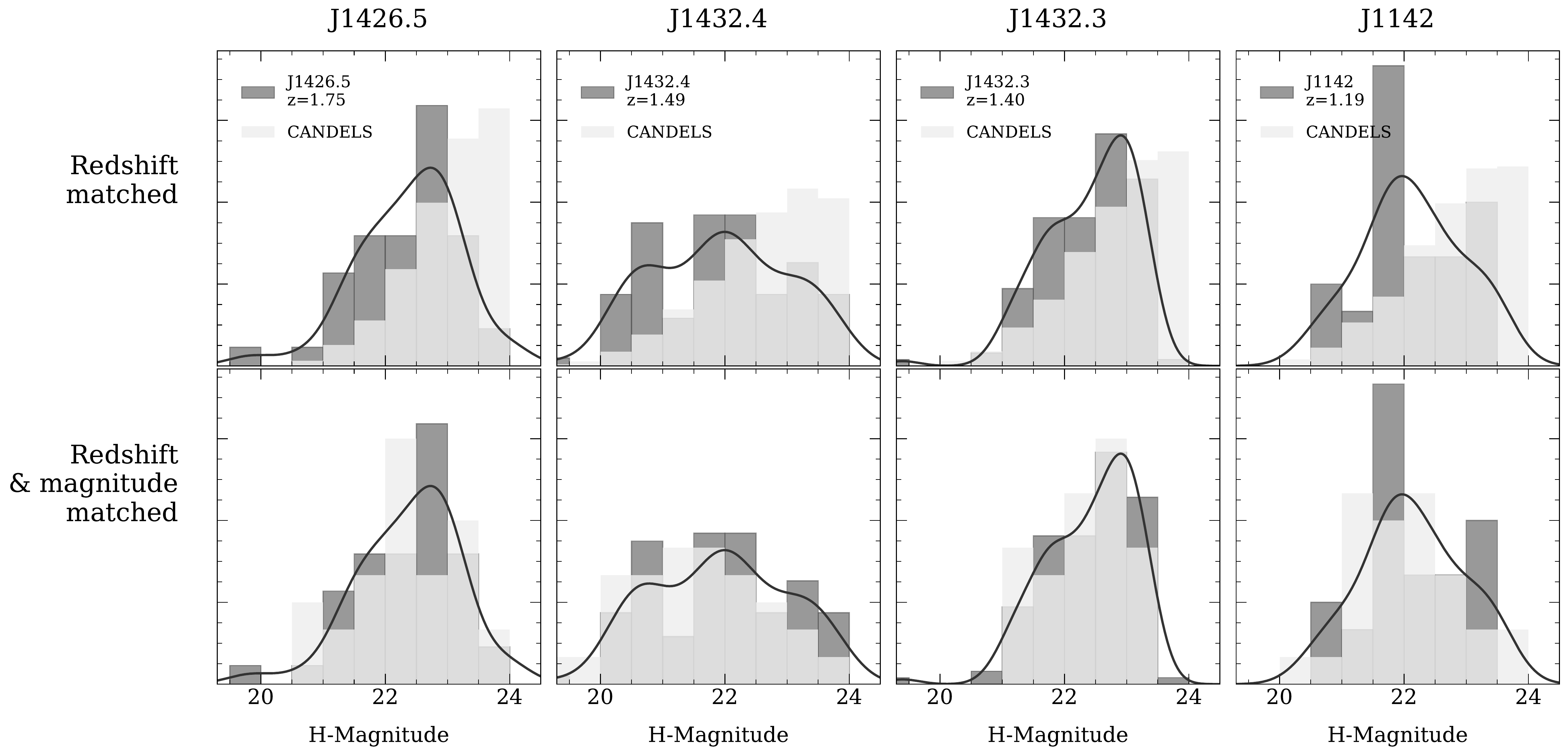}
\caption{\textbf{Top:} H-magnitude distribution of cluster galaxies (dark grey) and CANDELS field galaxies (light greys) redshift-matched within $\Delta z = 0.25$. The dark grey line shows a Gaussian KDE fit to each cluster distribution. A clear mass assembly bias is seen, with field galaxies being on average dimmer than cluster galaxies. \textbf{Bottom:} the same cluster H-magnitude distribution (dark grey), and a magnitude-matched CANDELS field distribution (light grey) for each cluster. 40,000 iterations of this magnitude-matching are performed to construct the Monte Carlo control ensemble.}
\label{fig:mcmc_all}
\end{figure*}

As described in Section \ref{sec:control_sample}, we need to construct control samples that are free of both redshift bias, and a mass assembly bias caused by a larger average mass of cluster galaxies. We construct control samples that match cluster redshift within $\Delta z = 0.25$. Using H-magnitude as a proxy for stellar mass, we then select control galaxies as to match the cluster magnitude distribution. To do so, we fit a Gaussian KDE to the cluster distribution, and select sample control galaxies to match the KDE. Figure \ref{fig:mcmc_all} shows the initial cluster and field distributions (top) and the resulting distributions after magnitude-matching (bottom) for all 4 clusters studied.

\section{Individual Morphology Distributions}\label{app:distributions}

\begin{figure*}[h!]
\centering
\includegraphics[width=1.3\textwidth, angle=270]{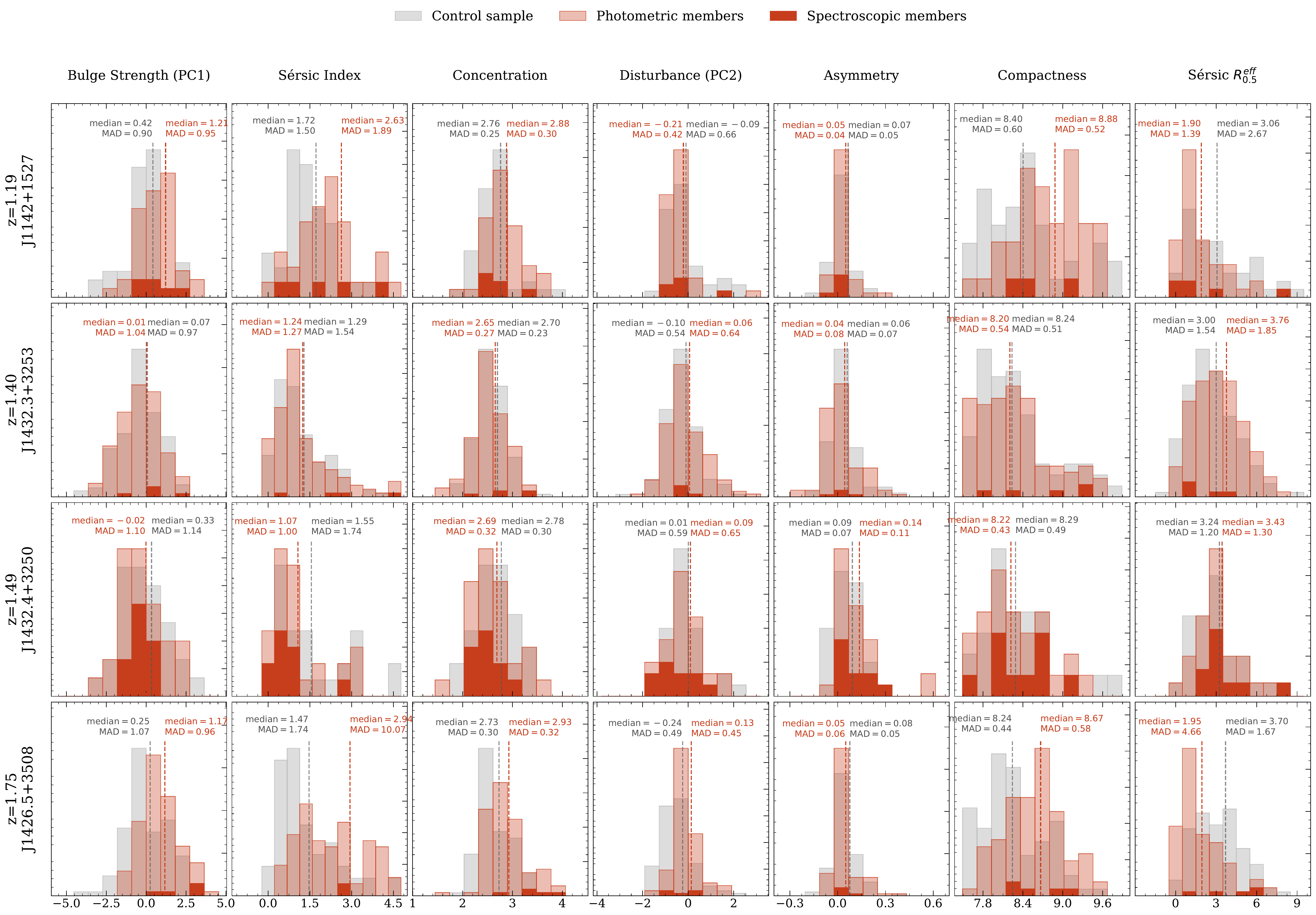}
\caption{Distribution of the morphological statistics outlined in Sec. \ref{sec:morph} for each cluster in Table \ref{tab:clusters}. Three distributions are shown: spectroscopic cluster members (dark red), photometric members (light red) and one iteration of magnitude- and redshift-matched control sample (grey). Dotted grey lines indicate the distribution median for cluster members (red) and the control sample (grey).}
\label{fig:distributions}
\end{figure*}

Morphology parameters outlined in Section \ref{sec:morph} are calculated for each galaxy in each cluster. As described in Sec. \ref{sec:mcmc}, the median of the distribution for each cluster is compared to the median of 40,000 Monte Carlo iterations of a field sample of equal size. Figure \ref{fig:distributions} shows the raw underlying distributions for a subset of parameters: Concentration, Asymmetry, S\'ersic index, S\'ersic effective radius, Bulge Strength and Disturbance. Cluster galaxies are shown in red, and the control sample in grey. 

It is clear from the plots that the distribution of morphologies in each cluster is wide, and no cluster is dominated by any one galaxy type. However, in the case of J1426.5 and J1142, it is clear the the distributions of bulge strength and S\'ersic index are offset, and comes from a qualitatively different population. On the other hand, the same distributions in J1432.3 and J1432.4 match the field galaxies closely.

\section{Median morphology values for field and cluster galaxies}\label{app:medians}

In Table \ref{tab:medians} we provide mean values for each morphological parameter measured in the field and cluster galaxies. For the field galaxies, we take the population median of the distribution of medians of 40,000 simulated control samples. The uncertainty on the population median is given by the 16$^{\textrm{th}}$ and 84$^{\textrm{th}}$ percentiles of the median distribution. For the cluster galaxies, since Monte Carlo sampling of the cluster sample is unfeasible, we took the cluster median, and bootstrapped the cluster sample 40,000 times with replacement to obtain the 16$^{\textrm{th}}$ and 84$^{\textrm{th}}$ percentiles of the cluster median distribution.

\renewcommand{\arraystretch}{1.2}
\begin{splitdeluxetable*}{r|DD|DDBr|DD|DD}
\tablecaption{Median morphological parameters from all cluster and field samples}
\tablehead{
\multicolumn1r{Parameter} & \multicolumn2c{J1142+1527} & \multicolumn2c{Control} & 
\multicolumn2c{J1432.3+3253} & \multicolumn2c{Control} & \multicolumn1r{Parameter} & \multicolumn2c{J1432.4+3250} & \multicolumn2c{Control} & \multicolumn2c{J1426.5+3508} & \multicolumn2c{Control}
}
\decimals
\startdata
Bulge Strength & $1.173^{+0.075}_{-0.090}$ & $0.449^{+0.144}_{-0.143}$ & $-0.020^{+0.300}_{-0.154}$ & $0.339^{+0.263}_{-0.272}$ & Bulge Strength & $0.013^{+0.231}_{-0.132}$ & $0.141^{+0.127}_{-0.119}$ & $1.214^{+0.186}_{-0.374}$ & $0.282^{+0.226}_{-0.231}$ \\
Compactness & $8.666^{+0.041}_{-0.075}$ & $8.228^{+0.071}_{-0.073}$ & $8.220^{+0.140}_{-0.099}$ & $8.319^{+0.148}_{-0.117}$ & Compactness & $8.203^{+0.057}_{-0.091}$ & $8.262^{+0.052}_{-0.072}$ & $8.882^{+0.184}_{-0.143}$ & $8.444^{+0.125}_{-0.115}$ \\
Disturbance & $0.133^{+0.074}_{-0.059}$ & $-0.192^{+0.065}_{-0.060}$ & $0.095^{+0.134}_{-0.255}$ & $-0.138^{+0.117}_{-0.115}$ & Disturbance & $0.065^{+0.078}_{-0.082}$ & $-0.110^{+0.065}_{-0.052}$ & $-0.210^{+0.057}_{-0.062}$ & $-0.042^{+0.116}_{-0.095}$ \\
Magnitude & $21.986^{+0.052}_{-0.148}$ & $21.937^{+0.151}_{-0.149}$ & $21.980^{+0.288}_{-0.122}$ & $22.120^{+0.196}_{-0.188}$ & Magnitude & $22.544^{+0.086}_{-0.097}$ & $22.537^{+0.092}_{-0.095}$ & $22.534^{+0.085}_{-0.284}$ & $22.455^{+0.161}_{-0.165}$ \\
$\textrm{S\'ersic Index}$ & $2.941^{+0.140}_{-0.256}$ & $1.527^{+0.184}_{-0.159}$ & $1.070^{+0.081}_{-0.184}$ & $1.514^{+0.347}_{-0.294}$ & $\textrm{S\'ersic Index}$ & $1.243^{+0.060}_{-0.073}$ & $1.294^{+0.149}_{-0.096}$ & $2.631^{+0.106}_{-0.217}$ & $1.609^{+0.332}_{-0.286}$ \\
Concentration & $2.927^{+0.047}_{-0.048}$ & $2.760^{+0.047}_{-0.040}$ & $2.686^{+0.085}_{-0.096}$ & $2.724^{+0.081}_{-0.069}$ & Concentration & $2.652^{+0.033}_{-0.038}$ & $2.679^{+0.031}_{-0.030}$ & $2.881^{+0.086}_{-0.049}$ & $2.717^{+0.070}_{-0.058}$ \\
$R_{0.5, \textrm{S\'ersic}}$ (kpc) & $1.951^{+0.251}_{-0.132}$ & $3.212^{+0.273}_{-0.245}$ & $3.432^{+0.137}_{-0.124}$ & $3.124^{+0.459}_{-0.414}$ & $R_{0.5, \textrm{S\'ersic}}$ (kpc) & $3.760^{+0.132}_{-0.244}$ & $3.041^{+0.184}_{-0.206}$ & $1.901^{+0.156}_{-0.043}$ & $3.023^{+0.407}_{-0.344}$ \\
Asymmetry & $0.050^{+0.005}_{-0.005}$ & $0.068^{+0.006}_{-0.006}$ & $0.136^{+0.023}_{-0.060}$ & $0.069^{+0.015}_{-0.011}$ & Asymmetry & $0.044^{+0.003}_{-0.009}$ & $0.065^{+0.006}_{-0.006}$ & $0.051^{+0.006}_{-0.009}$ & $0.065^{+0.013}_{-0.011}$ \\
$M_{20}$ & $-1.737^{+0.012}_{-0.020}$ & $-1.704^{+0.020}_{-0.021}$ & $-1.636^{+0.014}_{-0.036}$ & $-1.679^{+0.034}_{-0.038}$ & $M_{20}$ & $-1.613^{+0.013}_{-0.038}$ & $-1.654^{+0.019}_{-0.015}$ & $-1.803^{+0.026}_{-0.028}$ & $-1.658^{+0.038}_{-0.036}$ \\
Gini & $0.524^{+0.012}_{-0.005}$ & $0.493^{+0.006}_{-0.006}$ & $0.487^{+0.006}_{-0.009}$ & $0.490^{+0.011}_{-0.010}$ & Gini & $0.494^{+0.006}_{-0.006}$ & $0.486^{+0.005}_{-0.005}$ & $0.513^{+0.004}_{-0.008}$ & $0.491^{+0.009}_{-0.008}$ \\
\enddata\label{tab:medians}
\end{splitdeluxetable*}

\vfill
\end{document}